\documentclass[prm,reprint]{revtex4-2}
\usepackage{setspace, parskip, xcolor, breakcites, hyphenat, graphicx, epstopdf, amsmath, amssymb, amsfonts, array,
	verbatim, url, subcaption, booktabs, multirow, etoolbox,siunitx, etoolbox, tikz,stackengine, textcomp,
	gensymb, siunitx,tabularx,array}
\newcolumntype{H}{>{\setbox0=\hbox\bgroup}c<{\egroup}@{}}
\usepackage[toc,page,titletoc]{appendix}
\usepackage[inline]{enumitem}
\usepackage[version=4]{mhchem}
\usepackage[T1]{fontenc}
\usepackage[font=small,labelfont=bf]{caption}
\usetikzlibrary{shadings}
\definecolor{srm}{HTML}{034da1}
\definecolor{cof}{RGB}{219,144,71}
\definecolor{pur}{RGB}{186,146,162}
\definecolor{greeo}{RGB}{91,173,69}
\definecolor{greet}{RGB}{52,111,72}
\definecolor{mplc0}{HTML}{1f77b4}
\definecolor{mplc1}{HTML}{ff7f0e}
\definecolor{mplc2}{HTML}{2e7d32}
\definecolor{mplc3}{HTML}{d32f2f}
\definecolor{mplc4}{HTML}{9467bd}
\definecolor{mplc5}{HTML}{8c564b}
\definecolor{mplc6}{HTML}{e377c2}
\definecolor{mplc7}{HTML}{7f7f7f}
\definecolor{mplc8}{HTML}{bcbd22}
\definecolor{mplc9}{HTML}{17becf}
\definecolor{Astral}{HTML}{1F77B4}
\definecolor{BG80}{HTML}{37474f}
\definecolor{Green}{HTML}{4CA540}
\definecolor{Red}{HTML}{F44336}
\usepackage[colorlinks=true, linkcolor=mplc3, urlcolor=mplc3, filecolor=mplc0, citecolor=mplc2,
	pdfstartview=FitV, pdftitle={}, pdfauthor={}, pdfsubject={}, pdfkeywords={}, pdfpagemode={},
	bookmarksopen=true, breaklinks]{hyperref}

\usepackage[capitalise,nameinlink]{cleveref}
\crefname{supp}{Supplement}{Supplements}
\PassOptionsToPackage{hyphens}{url}

\newcommand{\figref}[1]{figure (\ref{#1})}
\newcommand{\eqnref}[1]{eqn\eqref{#1}}
\newcommand{\tabref}[1]{table (\ref{#1})}

\graphicspath{{./figures}}
\date{\today}

\makeatletter
\def\@email#1#2{%
	\endgroup
	\patchcmd{\titleblock@produce}
	{\frontmatter@RRAPformat}
	{\frontmatter@RRAPformat{\produce@RRAP{*#1\href{mailto:#2}{#2}}}\frontmatter@RRAPformat}
	{}{}
}%
\makeatother
\begin{document}

\title{Auditing Machine-Learning Models and Their Training Data with Explainability and First-Principles Verification: Application to Spin Hall Conductivity}
% \title{A Diagnostic Protocol for Element-as-Proxy Pathologies in Composition-Only Machine Learning: Application to Spin Hall Conductivity}
% \title{Interpretable Composition-Based Prediction of Spin Hall Conductivity Reveals Descriptor-Driven Bias in Data-Driven Discovery}
% \title{Accelerated Discovery of Efficient Spin-to-Charge Transfer Materials using Explainable Machine Learning}
% \title{Beyond Diamond: Interpretable Machine Learning Identifies Quantum Defect Host Materials}
% \title{A Generalizable Framework for Discovering Coherent Quantum Defect Hosts via Interpretable Machine Learning}
% \title{Composition-Only Machine Learning Predicts Coherent Quantum Defect Hosts}
% \title{Rashomon Ensembles for Interpretable Quantum Defect Host Discovery}
\author{Mohammed Mahshook}
\author{Rudra Banerjee*}\email{rudrab@srmist.edu.in}
\affiliation{Department of Physics and Nanotechnology, SRM Institute of Science and Technology, Kattankulathur, Tamil
	Nadu, 603203, India}

\begin{abstract}
	Machine-learning models for materials properties rest on two assumptions that standard validation never tests: that a
	model's features reflect the physics of the property rather than accidents of the training distribution, and that the
	training labels are themselves correct. We introduce a model-agnostic audit protocol for both, combining SHAP attribution,
	counterfactual partial-dependence analysis, and Rashomon-style cross-model verification, with every finding adjudicated by
	targeted density functional theory (DFT). Demonstrated on intrinsic spin Hall conductivity using a composition-only Random
	Forest, the model needs no relaxed crystal structure, reaching accuracy competitive with structure-aware graph networks
	while remaining applicable to the far larger space of compositions for which no structure has been computed. The
	\emph{model audit} reveals that the average $p$-valence descriptor becomes statistically entangled with Pt content --- a
	property of the learned representation rather than the physics; DFT confirms the consequence, a Pt-free compound
	(\ce{HgOsPb2}) whose true SHC is nearly four times the prediction. The \emph{data audit} exposes a thirtyfold error in the
	\ce{HfC} training label, inherited undetectably by every black-box model trained on the same data. The protocol audits a
	model and its training data for the cost of a few DFT calculations, wherever one element dominates the high-property regime.
\end{abstract}

\maketitle
\section{Introduction}
Machine-learning surrogates are now routinely used to screen materials for target properties at scales far beyond the reach of
first-principles calculation. Their value rests on a chain of trust that is rarely examined: that the model has learned the
physics rather than a statistical shortcut, and that the data it learned from is correct. Intrinsic spin Hall conductivity (SHC)
is an unusually demanding test of that trust. The spin Hall effect (SHE) - a transverse spin current generated by a longitudinal
charge current - underpins spin-to-charge interconversion in modern spintronics\cite{sinova2015spin}, with SHC as its principal
figure of merit; and first-principles work has revealed SHC values an order of magnitude above the canonical Pt and $\beta$-Ta
benchmarks\cite{sagasta2016tuning, sagasta2018unveiling} in A15 superconductors\cite{derunova2019giant, sattigeri2024dirac},
Dirac semimetals and dichalcogenides\cite{xu2020high, zhou2018intrinsic}, noncollinear
antiferromagnets\cite{zhu2024crystal}, and nonmagnetic crystals with mirror-symmetry-protected nodal
lines\cite{zhang2021different}. In all cases the SHC arises from a specific interplay of strong spin-orbit coupling (SOC) and
electronic-structure features near $E_F$, accessible only through Kubo-Wannier integration on dense
$k$-grids\cite{ado2024kubo, qiao2018calculation} at a cost of $\sim 10^3$ core-hours per fully-relaxed compound. Machine
learning has been deployed against this bottleneck: graph networks\cite{zhao2024accelerating} and
transformers\cite{zhang2025predicting} predict SHC from relaxed structures, while screening campaigns have targeted
Heusler\cite{ji2022spin} and A15\cite{derunova2019giant, sattigeri2024dirac} families, and interpretable low-dimensional models
have been demonstrated for related transport properties\cite{kipp2024symmetry, ouyang2018sisso}.

These models share two untested assumptions. The first is that the features a model uses reflect the physics of the property
rather than statistical accidents of the training distribution; the second is that the training labels themselves are correct.
Both assumptions are consequential - a model that has learned an elemental shortcut will silently fail on chemistries where the
shortcut breaks, and a model trained on erroneous labels will reproduce the errors - yet neither is routinely examined, because
black-box predictors offer no mechanism for examining them. Feature attribution methods such as
SHAP\cite{lundberg2017unified} are increasingly used in materials informatics\cite{pilania2021machine}, but typically stop at
importance rankings; their application to SHC has not been reported, and their use to \emph{audit} a model and its training
data, with first-principles verification of what the audit finds, has to our knowledge not been demonstrated for any transport
property.

This work introduces such an audit. Our central contribution is a model-agnostic protocol - global SHAP attribution,
counterfactual partial-dependence analysis, and Rashomon-style cross-model
verification\cite{breiman2001statistical, semenova2022existence} - that interrogates a composition-based property model on two
fronts: whether its learned representation harbors \emph{element-as-proxy} dependencies (a model audit), and whether
individual training labels are consistent with what the model and the physics imply (a data audit). Each audit finding is then
subjected to independent density functional theory (DFT) calculation, which either grounds it in physics or refutes it. The
protocol is not specific to any model class - it applies to any predictor whose inputs admit attribution, including
structure-aware graph networks - and it does not depend on the particular dependencies it happens to find: the demonstrations
below could be replaced by others without altering the method.

We demonstrate the audit on intrinsic SHC. As its basis, a Random Forest trained on a 211-dimensional descriptor, organized in
three physically motivated tiers (atomic SOC strength, orbital-resolved hybridization, Fermi-level electronic-structure
context), attains a test MAE of $114.5~(\hbar/e)(S/cm)$ on the dataset of
Zhao~\textit{et~al.}\cite{zhao2024accelerating} - below both the structure-aware CGCNN ($126.7$) and
Res-CGCNN ($118.7$) despite using no structural input. We treat this not as a state-of-the-art claim but as the license to
interrogate the model.

The model audit finds that, within this dataset, the average $p$-valence orbital descriptor becomes statistically entangled with
the presence of Pt - a representational dependency, not a physical law about SHC - consistent across the Random Forest and an
independently trained Gaussian Process. Independent DFT confirms the predicted consequence: a Pt-free compound, \ce{HgOsPb2},
whose true SHC is nearly four times the model's prediction, the under-flagging the audit anticipates. The data audit, triggered
by a conspicuous disagreement between the model and a training label, identifies a probable error in the label of \ce{HfC},
which our independent calculation exceeds by a factor of $\sim 30$ - an error that every black-box model trained on the same
dataset inherits without any means of detecting it.

The remainder of the paper is organized as follows. We first establish the predictive model and its descriptor, then apply the
audit protocol - global attribution, counterfactual analysis, and cross-model verification - to diagnose the Pt-orbital
entanglement and test whether it has exploitable screening consequences. We then deploy the model across $\sim 40000$ Materials
Project compositions and, finally, subject the audit's central findings to first-principles verification, confirming both the
model audit and the data audit on independent calculations.

\section{Methods}

\subsection{Dataset Construction and Machine Learning Framework}

We use the DFT dataset of 9249 nonmagnetic compounds reported by Zhao~\textit{et~al.}\cite{zhao2024accelerating}, in which the
training label is the maximum absolute component of the SHC tensor evaluated by the Kubo-Wannier method. Two pre-processing
steps were applied prior to training. First, the distribution of SHC values is severely right-skewed and strongly biased toward
the low-SHC region: more than half of the training entries lie below $200~(\hbar/e)(S/cm)$, while the high-SHC tail extends
beyond $2000~(\hbar/e)(S/cm)$ with only a small number of supporting examples. Training on this raw distribution produces models
that systematically under-predict the high-SHC tail, since the loss landscape rewards accuracy on the densely-populated
low-SHC region at the expense of the sparse but materials-relevant high-SHC end. To mitigate this, we applied a Box-Cox
transformation to the target,
\begin{equation}
	f_\lambda(y) =
	\begin{cases}
		\dfrac{y^\lambda - 1}{\lambda} & \text{if } \lambda \neq 0 \\
		\ln(y)                         & \text{if } \lambda = 0
	\end{cases}
\end{equation}
where $y$ is the maximum-absolute-component SHC (non-negative by construction) and $\lambda$ is the transformation parameter,
optimized by maximum-likelihood estimation. The dataset contained only two entries with zero reported SHC; since the
transformation requires strictly positive values, these entries were removed rather than shifted (see below).
The transformation compresses the long high-SHC tail and renders the distribution approximately Gaussian, satisfying the
homoscedasticity assumptions of the regression algorithms used. Critically, no instances are discarded as outliers - the
materially-interesting high-SHC examples remain in the training set but are reweighted by the transformation. We compute all
reported error metrics in the original SHC units after applying the inverse Box-Cox transformation; the qualitative effect of
the transformation is presented in the Results section.

Second, although the SHC values across polymorphs of the same composition are nearly identical for most entries (average
standard deviation $\approx 259~(\hbar/e)(S/cm)$), polymorphs all map to the same vector in compositional feature space. Without
intervention this introduces label noise and, more seriously, leakage between train and test splits. We therefore reduced
polymorphs to a single composition prior to splitting using the following rule:
\begin{enumerate*}[label=(\roman*)]
	\item Materials are mapped to their Materials Project entries.
	\item If the entry exists in Materials Project, the most stable polymorph (lowest energy above the convex hull) and its
	      corresponding SHC value are retained.
	\item If no Materials Project match exists but multiple polymorphs are present in the dataset, the composition is removed
	      entirely to avoid arbitrary selection.
\end{enumerate*}
This rule reduced the dataset from 9249 to 7515 compositions (an attrition of 18.7\%); two further entries with zero reported
SHC were removed prior to the Box-Cox transformation, leaving 7513. The reduced dataset was split into training, validation, and
test partitions in a 7:1:2 ratio, yielding a held-out test set of 1503 compositions. The validation partition was used for
hyperparameter selection; the test partition was held out and used only for the final reported metrics.

The compositional descriptor was designed in three physically motivated tiers, each capturing one of the contributions to
intrinsic SHE. The first tier, encoding \textit{atomic SOC strength}, consists of 118 element-fraction indicators covering the
periodic table; these scale directly with atomic number and provide the model with access to the dominant atomic ingredient of
SOC. The second tier, encoding \textit{orbital-resolved hybridization}, comprises 20 descriptors - fractional valence-electron
counts ($s$-, $p$-, $d$-, $f$-fractions), stoichiometric $L^p$ norms ($p \in \{0,2,3,5,7,10\}$), HOMO and LUMO energies with
their atomic-orbital gap, the band center, and HOMO/LUMO orbital character ($s$, $p$, $d$) - which proxy the $p$-$d$ and $d$-$f$
band-crossing physics underlying intrinsic Berry curvature. The third tier, encoding \textit{Fermi-level electronic-structure
	context}, comprises 72 Magpie\cite{ward2016general} valence- and unfilled-electron statistics together with ground-state
band-gap and magnetic-moment statistics, augmented by one engineered descriptor, $s_{\min}$, which measures the agreement
between the Magpie minimum-band-gap estimate and the atomic-orbital-derived gap ($E_{\text{HOMO}} - E_{\text{LUMO}}$) and was
introduced after observing that disagreement between these two estimates correlates with the presence of strong band
hybridization in the training set. The total descriptor dimensionality is 211. Featurization was performed with the matminer
library\cite{ward2018matminer}. The complete feature list, all hyperparameters, and the construction rationale for $s_{\min}$
are provided in the supplementary information.

We benchmarked three model classes - Random Forest (RF), Extreme Gradient Boosting (XGB), and Kernel Ridge Regression (KRR) -
against three feature sets: the proposed 211-feature set, the Magpie\cite{ward2016general} set, and the
Meredig\cite{meredig2014combinatorial} set. The full nine-way comparison is presented in the Results section; the RF regressor trained on
the proposed feature set achieved the best test performance and was adopted as the primary predictive model. RF also offers a
practical advantage for the audit itself: exact TreeSHAP attribution\cite{lundberg2020local} is available for tree ensembles,
making the attribution step of the protocol fast and free of sampling approximation.

To complement the point predictions of the Random Forest, we independently trained a Gaussian Process Regression (GPR) model on
the same feature set and target. The GPR predicts a distribution
\begin{equation}
	y_* \mid \mathbf{x}_*, X, \mathbf{y} \sim \mathcal{N}\bigl(\mu(\mathbf{x}_*), \sigma^2(\mathbf{x}_*)\bigr),
\end{equation}
in which $\mu(\mathbf{x}_*)$ is taken as the GPR's point prediction and $\sigma^2(\mathbf{x}_*)$ is the per-instance predictive
variance, used as a screening filter. We employed an RBF kernel with three optimized hyperparameters.
%\rev{[Removed the details of
%			the hyperparameters]}
%: RBF variance
%$221.93~(\hbar/e)^2(S/cm)^2$, RBF lengthscale $12.67$, and Gaussian noise variance $34.25~(\hbar/e)^2(S/cm)^2$.
The GPR achieves
a test MAE of $127.16~(\hbar/e)(S/cm)$, which is poorer than the RF in absolute terms but qualitatively consistent. Because RF
and GPR have very different inductive biases - RF builds piecewise-constant predictions from axis-aligned splits, while GPR
imposes a smoothness prior in the RBF feature similarity - agreement between the two models on qualitative trends and bias
direction is more meaningful than agreement on point predictions. We exploit this Rashomon-style consistency in the bias
analysis below.

\subsection{Exploratory probe of screening consequences}

To probe whether the diagnosed Pt-orbital entanglement has a measurable consequence for screening, we evaluated a bias-aware
reranking of the held-out test set, structurally analogous to post-processing corrections in fairness-aware supervised
learning\cite{hardt2016equality, rosenbaum1983propensity}. We emphasize that this is a diagnostic probe, not a proposed
screening method: the correction strength is swept over a range and evaluated on the same set, which can reveal whether an
exploitable signal exists but cannot yield a generalizable correction (this limitation is discussed with the results). For each
composition $i$ we computed an adjusted ranking score
\begin{align}
	\begin{split}
		s_{\text{adj}}(i) & = \alpha(i) \cdot s_{\text{RF}}(i) \\%, \quad
		\alpha(i)         & =
		\begin{cases}
			\alpha_0 & \forall~i  \begin{cases}
				                      \ni     & \text{a heavy element} \\
				                      \not\ni & \ce{Pt}
			                      \end{cases} \\
			1        & \text{otherwise}
		\end{cases}
	\end{split}
	\label{eq:bias_aware}
\end{align}
where $s_{\text{RF}}(i)$ is the RF-predicted SHC, the heavy-element set is
$\{\text{Ta}, \text{W}, \text{Ir}, \text{Au}, \text{Bi}, \text{Hg}, \text{Hf}, \text{Re}, \text{Os}, \text{Pd}\}$, and
$\alpha_0$ is a multiplicative correction factor swept over $\{1.0, 1.2, 1.3, 1.4, 1.5, 1.6, 1.8, 2.0\}$ ($\alpha_0 = 1.0$
recovering naive ranking).

We measured performance using recall at $k$,
\begin{equation}
	R_k = \frac{|\{i : i \in \text{top-}k(s_{\text{adj}}) \,\wedge\, y_i \geq y_{\text{high}}\}|}{|\{i : y_i \geq y_{\text{high}}\}|}
\end{equation}
for $k \in \{20, 50, 100, 150, 200\}$ and $y_{\text{high}} \in \{500, 1000, 1500, 2000\}~(\hbar/e)(S/cm)$, and additionally
tested an additive correction $s_{\text{adj}}(i) = s_{\text{RF}}(i) + \beta \cdot \mathbb{1}[\text{heavy, non-Pt}]$ for
$\beta \in \{100, \dots, 700\}~(\hbar/e)(S/cm)$ as a sensitivity check.

\subsection{First-principles Calculations}
To validate the machine learning predictions and investigate the electronic origins of the spin Hall effect, we performed
first-principles density functional theory (DFT) calculations using the Quantum ESPRESSO package\cite{giannozzi2009j,
	giannozzi2017ukbenli}. The electronic exchange-correlation interactions were treated within the generalized gradient
approximation (GGA) using the Perdew-Burke-Ernzerhof (PBE) functional. To account for relativistic effects essential for the
spin Hall effect, we employed fully relativistic pseudopotentials with spin-orbit coupling (SOC) enabled for all calculations.

The crystal structures were fully relaxed until the forces on each atom were less than $10^{-4}$~Ry/au. For the electronic
structure and subsequent transport properties, the intrinsic SHC was computed using the Kubo-Greenwood
formalism\cite{ado2024kubo} via Wannier interpolation. The intrinsic SHC ($\sigma_{xy}^z$) is given by
\begin{equation}
	\sigma_{xy}^z = \frac{e}{\hbar} \int_{BZ} \frac{d^3k}{(2\pi)^3} \sum_n f_{nk}\, \Omega_{n,xy}^z(\mathbf{k}),
\end{equation}
where $f_{nk}$ is the Fermi-Dirac distribution for the $n^{\text{th}}$ band at wavevector $\mathbf{k}$, and
$\Omega_{n,xy}^z(\mathbf{k})$ is the spin Berry curvature,
\begin{equation}
	\Omega_{n,xy}^z(\mathbf{k}) = -2\,\text{Im}\sum_{m \neq n}
	\frac{\langle u_{nk} | \hat{j}_x^z | u_{mk} \rangle \langle u_{mk} | \hat{v}_y | u_{nk} \rangle}{(E_{nk} - E_{mk})^2},
\end{equation}
with $\hat{j}_x^z = \frac{\hbar}{4} \{ \hat{\sigma}_z, \hat{v}_x \}$ the spin-current operator. We constructed
maximally-localized Wannier functions (MLWFs) to obtain a tight-binding Hamiltonian, and integrated over the Brillouin zone on a
dense $200 \times 200 \times 200$ $k$-point grid to ensure convergence of the Berry-curvature hotspots. Linear-response
properties were evaluated using the WannierBerri code\cite{tsirkin2021high}, following the SHC methodology established by
Qiao~\textit{et~al.}\cite{qiao2018calculation}.

\section{Results}

We begin by presenting the qualitative effect of the pre-processing steps described in Methods, and then turn to model
performance, feature attribution, and large-scale screening.

\subsection{Effect of Box-Cox pre-processing}

\begin{figure*}[ht]
	\centering
	\begin{subfigure}{0.48\textwidth}
		\includegraphics[width=\textwidth]{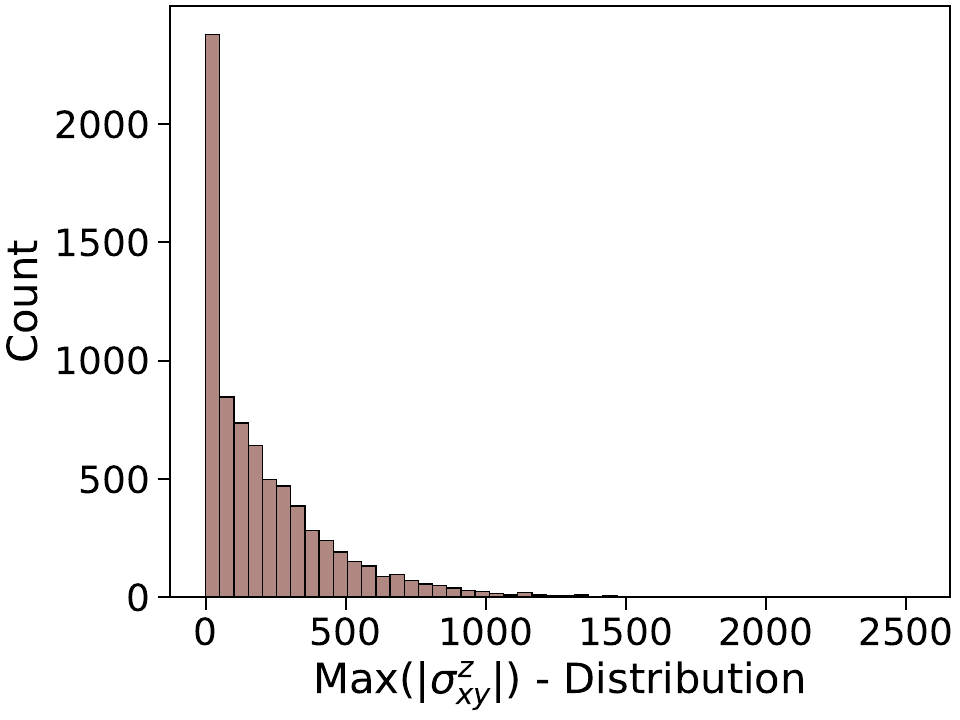}
		\caption{}
		\label{fig:raw_counts}
	\end{subfigure}
	\begin{subfigure}{0.48\textwidth}
		\includegraphics[width=\textwidth]{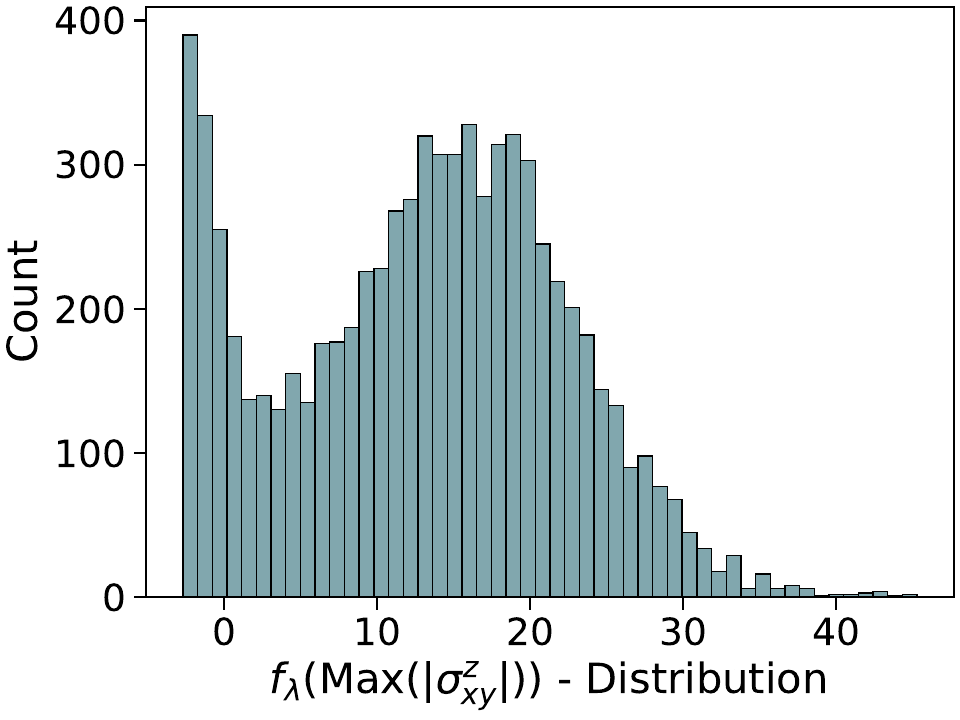}
		\caption{}
		\label{fig:bx_counts}
	\end{subfigure}

	\caption{The distribution of SHC values across the polymorph-reduced dataset is severely right-skewed
		(\subref{fig:raw_counts}); a model trained on the raw distribution inherits this bias. Applying a Box-Cox transformation
		yields an approximately normal target distribution (\subref{fig:bx_counts}) suitable for homoscedastic regression; only
		the two zero-valued entries are excluded, and no instance is discarded as an outlier.}
	\label{fig:distribution}
\end{figure*}

The raw distribution of SHC values across the polymorph-reduced dataset (7513 compositions) is shown in \figref{fig:raw_counts}.
More than half the entries lie below $200~(\hbar/e)(S/cm)$, with a long, sparsely-populated tail extending past
$2000~(\hbar/e)(S/cm)$. A regressor trained directly on this distribution would inherit a systematic bias toward under-prediction
in the high-SHC region - precisely the regime of materials-discovery interest. The Box-Cox-transformed distribution
(\figref{fig:bx_counts}) is approximately Gaussian and compresses the high-SHC tail into a tractable dynamic range; only two
entries with zero reported SHC are excluded (the transformation requires strictly positive values), and no instance is discarded
as an outlier. We use the transformed target for all model fitting and report all final error metrics in the original SHC units
after the inverse transformation.

\subsection{Model performance establishes a basis for interrogation}

\begin{figure}[ht]
	\centering
	\includegraphics[width=0.9\columnwidth]{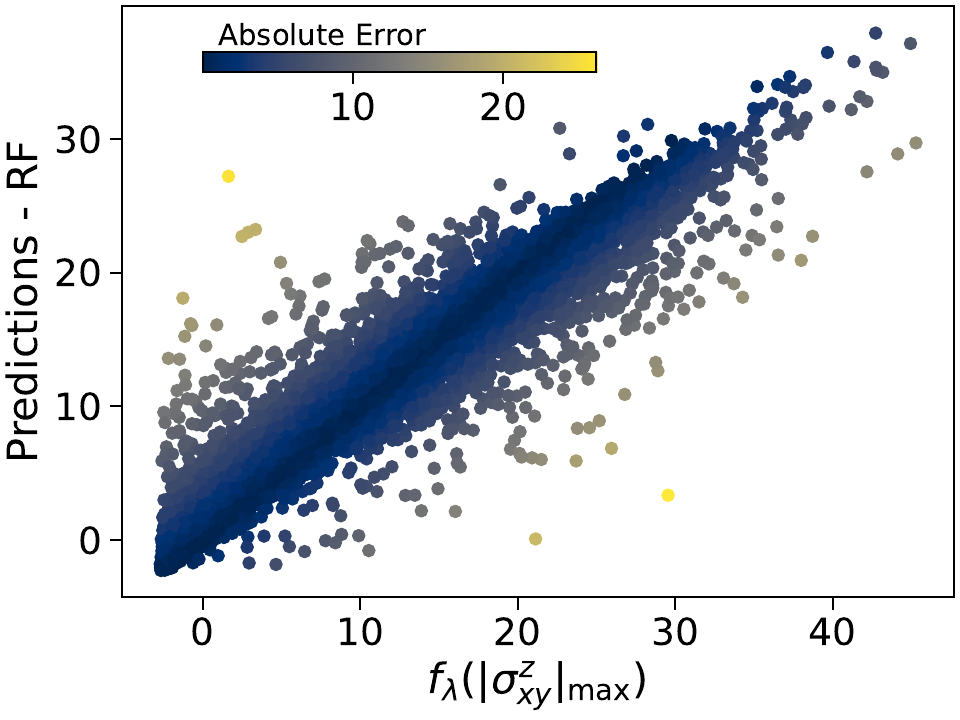}
	\caption{Cross-validated predictions versus true SHC values for the Random Forest model. The target is shown in
		Box-Cox-transformed units to eliminate the bias of the raw skewed distribution.}
	\label{fig:cv}
\end{figure}

Before interrogating the learned representation, we confirm that the composition-only model is accurate enough that its internal
structure is worth analyzing. On the fixed train/test split used throughout, the Random Forest regressor with the proposed
211-feature descriptor achieves a mean absolute error of $114.5~(\hbar/e)(S/cm)$; on the same dataset
Zhao~\textit{et~al.}\cite{zhao2024accelerating} report $126.7$ for CGCNN and $118.7~(\hbar/e)(S/cm)$ for the structure-aware
Res-CGCNN. The composition-only model is therefore competitive with both structure-aware baselines despite using no structural
input, though we do not regard the margin as the contribution of this work. As an independent robustness check, repeating the
fit over eight randomized splits gives a mean of $115.4 \pm 3.6~(\hbar/e)(S/cm)$, confirming that the reported performance is not
an artifact of the chosen split (supplementary information). The independently trained Gaussian Process attains a test MAE of
$127.16~(\hbar/e)(S/cm)$, comparable to CGCNN. Cross-validation across folds (\figref{fig:cv}) shows how the predictions fare in
boxcox-transformed units.
%no evidence of overfitting.

\begin{figure*}[ht]
	\begin{subfigure}{0.3\textwidth}
		\includegraphics[width=\textwidth]{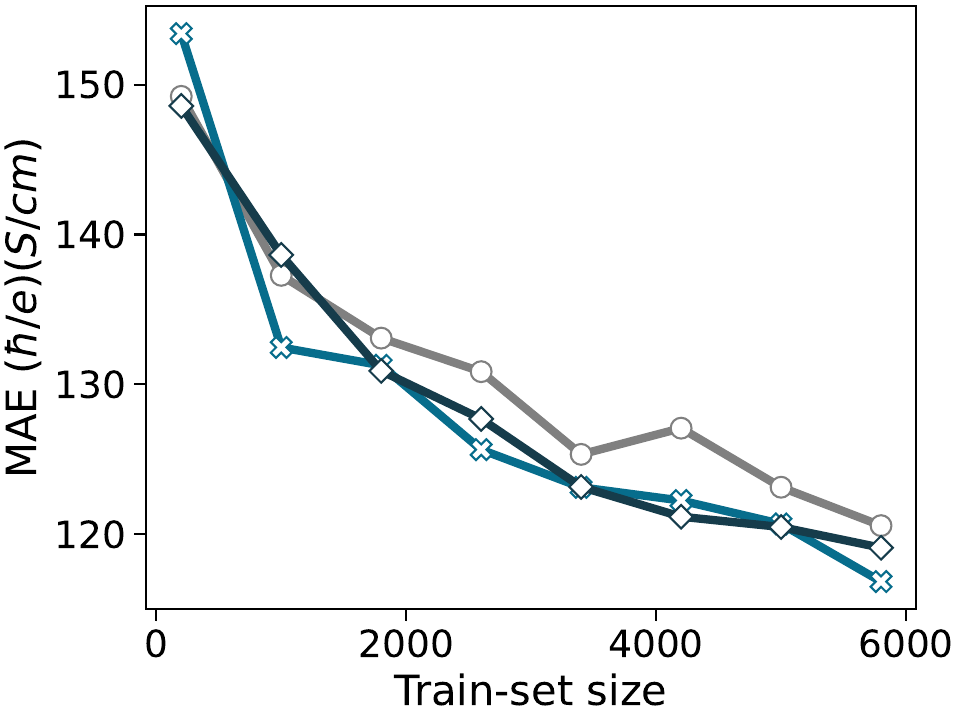}
		\caption{}
		\label{fig:rf}
	\end{subfigure}
	\begin{subfigure}{0.3\textwidth}
		\includegraphics[width=\textwidth]{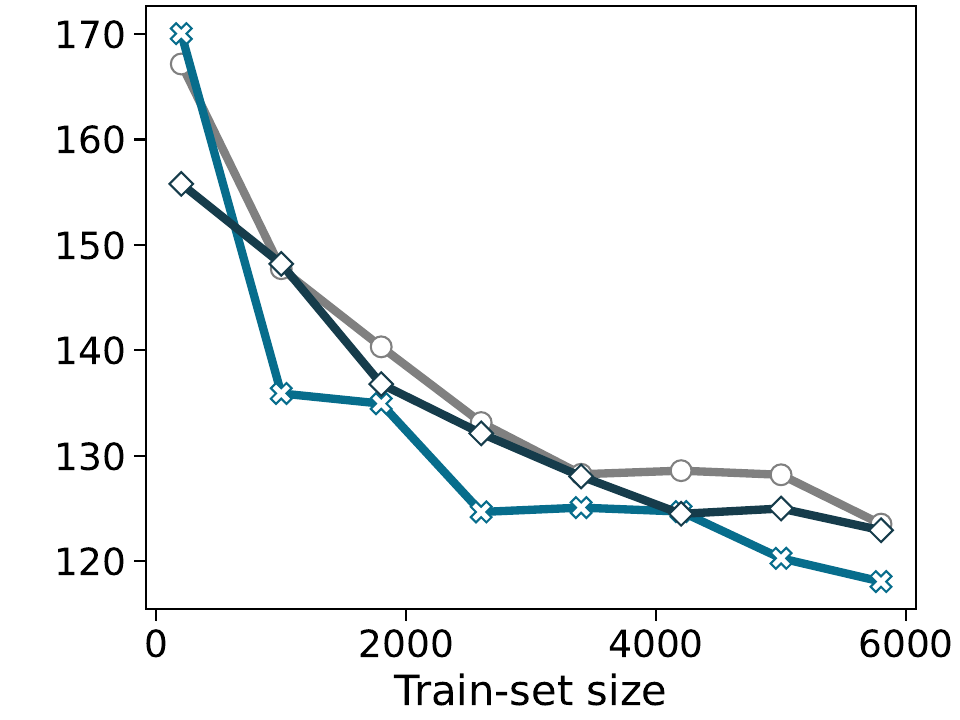}
		\caption{}
		\label{fig:xgb}
	\end{subfigure}
	\begin{subfigure}{0.3\textwidth}
		\includegraphics[width=\textwidth]{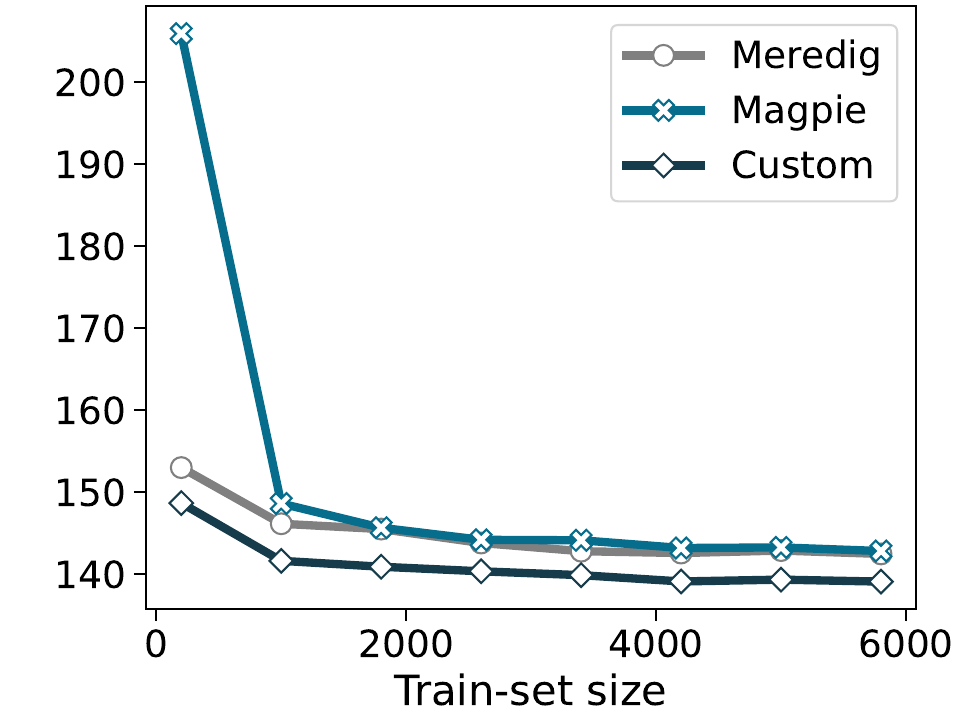}
		\caption{}
		\label{fig:krr}
	\end{subfigure}
	\caption{Benchmark of the proposed feature-importance-reduced 211-feature descriptor against the
		Magpie\cite{ward2016general} and Meredig\cite{meredig2014combinatorial} descriptor sets, across three model classes:
		Random Forest (\subref{fig:rf}), Extreme Gradient Boosting (\subref{fig:xgb}), and Kernel Ridge Regression
		(\subref{fig:krr}). The proposed set matches or modestly outperforms Magpie in every case and substantially outperforms
		Meredig throughout. The reduction from the 403-dimensional combined space to the 211-dimensional proposed space incurs no
		statistically significant change in test error ($115.4 \pm 3.6$ vs $116.8 \pm 2.9~(\hbar/e)(S/cm)$ across eight splits;
		see supplementary information), so the reduced space is preferred for its lower dimensionality, efficiency, and
		interpretability rather than for any accuracy gain.}
	\label{fig:scores}
\end{figure*}

The proposed 211-feature descriptor is itself the product of a feature-importance-guided reduction. A
general-purpose compositional feature set derived from the chemical and electronic environment of a composition (see
supplementary information for more details) is introduced and
benchmarked against the standard Meredig\cite{meredig2014combinatorial} and Magpie\cite{ward2016general} sets. The full nine-way
benchmark across three model classes and three feature sets is shown in \figref{fig:scores}. The proposed
descriptor matches or modestly outperforms the Magpie\cite{ward2016general} set across all three model classes and
substantially outperforms the Meredig\cite{meredig2014combinatorial} set; RF yields the lowest test error, XGB is close behind, and KRR
is uniformly weakest, consistent with its sensitivity to feature scaling in high dimensions. We adopt RF as the primary model on
this basis. We then identified - by probing model feature importances - the subset of Magpie
features that most strongly influence SHC prediction, and augmented the baseline with only that subset. The result is a
211-dimensional space that nearly halves the 403-dimensional full combined space while preserving accuracy: across eight
randomized splits the reduced space achieves $115.4 \pm 3.6~(\hbar/e)(S/cm)$ against $116.8 \pm 2.9$ for the full space, a
difference well within one standard deviation and not statistically significant ($p \approx 0.4$; supplementary information).
The reduction is therefore justified not by an accuracy gain but by parity at half the dimensionality - with the attendant
benefits of faster inference, reduced preprocessing, and, most importantly for the audit that follows, a descriptor in which
every feature maps directly to a chemical or electronic quantity.

We report MAE as the primary metric and note that $R^2$ is unreliable on this dataset. The RF $R^2$ is $0.62$, marginally above
the $0.60$ of Zhao~\textit{et~al.}\cite{zhao2024accelerating} and below the $0.75$ reported for the SHCTransformer of
Zhang~\textit{et~al.}\cite{zhang2025predicting}. As detailed in the supplementary information, the severe target skewness and
Box-Cox transformation make $R^2$ sensitive to fold composition and inverse-transform domain - the same RF model can be tuned
to report $R^2 \approx 0.7$ with no improvement in held-out MAE - so we do not rely on it for comparison.

\subsection{Global feature attribution}

We next examine which descriptors drive the model's predictions, using SHAP attribution on the trained Random Forest. Throughout
this analysis we distinguish carefully between two interpretations of any attribution: that a feature is a genuine physical
determinant of SHC, and that a feature is a statistical correlate of high SHC within this particular training distribution.
SHAP reports the latter; it cannot, on its own, establish the former. We flag below where the two interpretations diverge.

\begin{figure*}[ht]
	\begin{subfigure}{0.45\textwidth}
		\includegraphics[width=\textwidth]{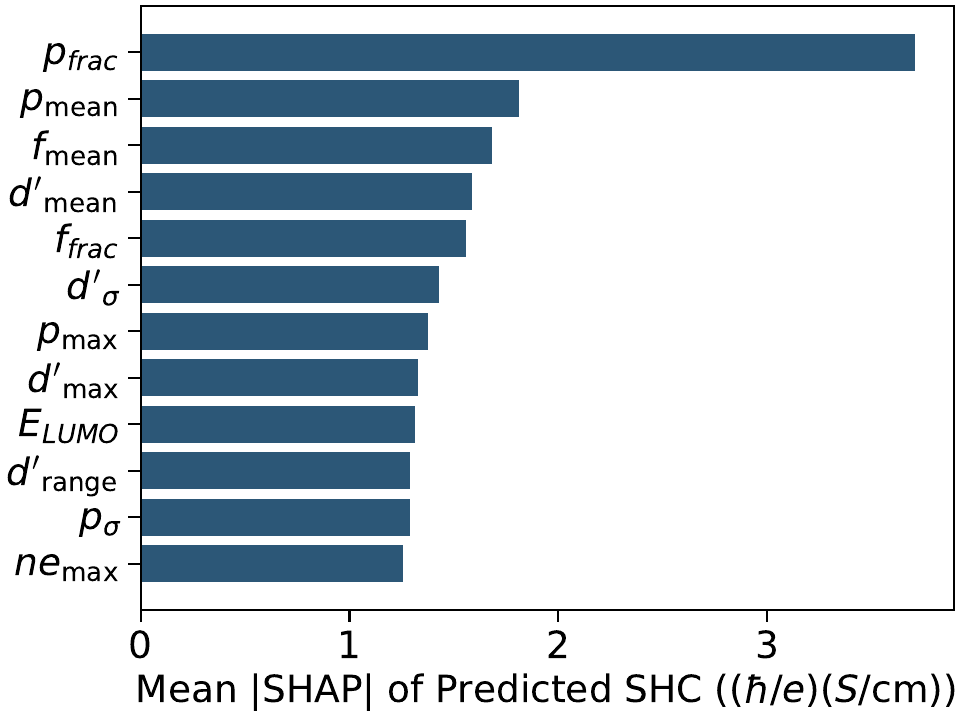}
		\caption{}
		\label{fig:fi}
	\end{subfigure}
	\begin{subfigure}{0.45\textwidth}
		\includegraphics[width=\textwidth]{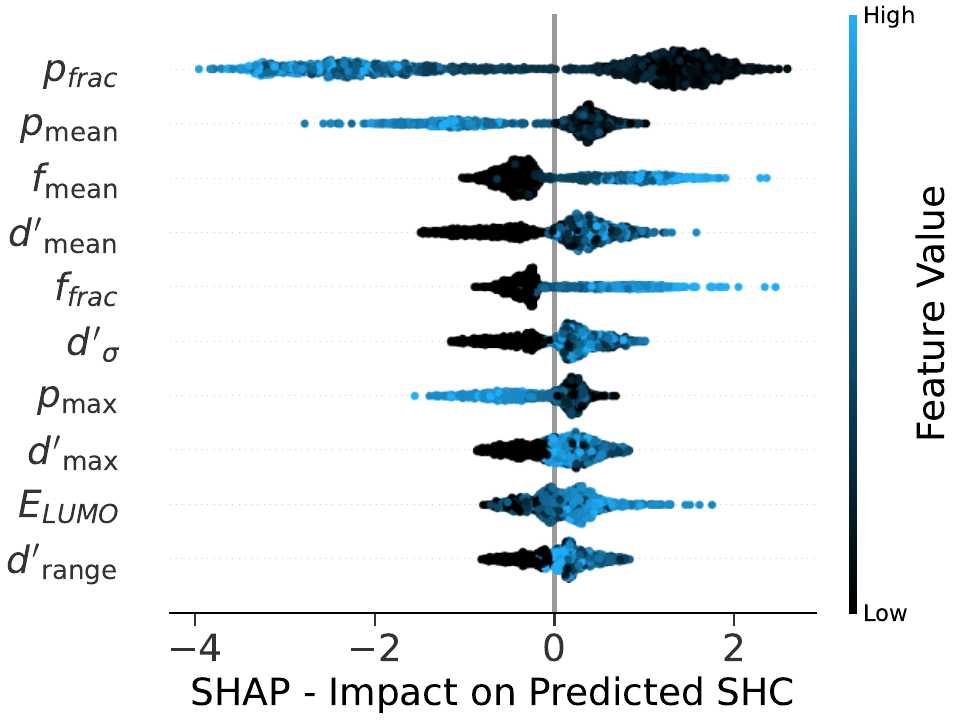}
		\caption{}
		\label{fig:beeswarm}
	\end{subfigure}

	\caption{SHAP analysis of the trained Random Forest model. Figure (\subref{fig:fi}) shows the absolute SHAP values averaged
		across the dataset, indicating the global feature importance of the most prominent descriptors. Figure
		(\subref{fig:beeswarm}) shows the feature-effect distribution of these prominent descriptors sampled across 200 random
		instances, highlighting how each feature influences the predicted Box-Cox-transformed SHC. Because Shapley values are
		additive in the model's prediction units, this additivity is lost under inverse-Box-Cox transformation; figure
		(\subref{fig:fi}) is shown in $(\hbar/e)(S/cm)$ for clarity, while figure (\subref{fig:beeswarm}) retains
		$f_\lambda(\sigma_{xy}^z)$ units.}
	\label{fig:shap}
\end{figure*}

The mean absolute SHAP values (\figref{fig:fi}) show that a small subset of descriptors dominates the prediction. Valence-electron
attributes populate this subset: notably, low $p$-valence-electron content together with unfilled $d$ and filled
$f$-valence electrons contribute significantly towards higher SHC values.
%[comment: This sentence was for an older version of shap plot, La
%			and Zr are longer identified as prominent featires]
%elemental
%presence indicators (notably La and Zr), averaged orbital descriptors ($\langle p \rangle$, $\langle s \rangle$), and
Further, electronic-structure proxies (the band center $E_C$, the atomic-orbital gap $E_{AO}$, and the LUMO energy) show positive
attribution. To verify that
these rankings are not artifacts of a particular split or hyperparameter choice, we repeated the attribution analysis across
20 bootstrap resamples of the training set; the leading features recur in the large majority of resamples, and
$p_{\text{frac}}$ - the descriptor central to the Pt-related entanglement reported below - is retained in the top ranks in
every resample (supplementary information).

The prominence of heavy elements (\ce{Pt, Bi, Ta, Os, Ir, W, Re}) is consistent with the
physics of intrinsic SHE, since SOC strength scales
steeply with atomic number. This is a case where the statistical correlate and the physical determinant plausibly coincide: the
model has inferred SOC-driven trends from composition alone. But this isn't consistent, throughout the feature set, statistical
correlates often diverge from meaningful physics.
%The high ranking of La is the first sign that the two
%interpretations can also diverge - La is over-represented among high-SHC entries in the Zhao~\textit{et~al.} dataset, so its
%attribution reflects dataset prevalence at least as much as any universal role as an SHC enhancer.
We return to a sharper instance of this divergence, involving Pt, in the next subsection.

The orbital descriptors ($\langle p \rangle$, $\langle d \rangle$, $\langle f \rangle$ and their fractional
counterparts) also rank highly, consistent with the understanding that intrinsic SHE arises from band crossings of specific
orbital character, particularly $p$-$d$ and $d$-$f$ hybridizations\cite{derunova2019giant, sattigeri2024dirac}. Several of these
features contribute bidirectionally - both low and high values of $p_{\text{frac}}$, $d_{\text{frac}}$, and $f_{\text{frac}}$
can shift the prediction in either direction depending on chemical context - reflecting the cooperative, nonlinear nature of
multi-orbital interactions and motivating the use of nonparametric ensembles. Electronic descriptors such as the LUMO energy and
band center carry substantial SHAP spread, indicating that band-alignment effects are measurable even at the elemental
approximation.

\subsection{Local explanations reveal a Pt-orbital entanglement in the learned representation}

\begin{figure*}[ht]
	\centering
	\begin{subfigure}{0.49\textwidth}
		\includegraphics[width=\textwidth]{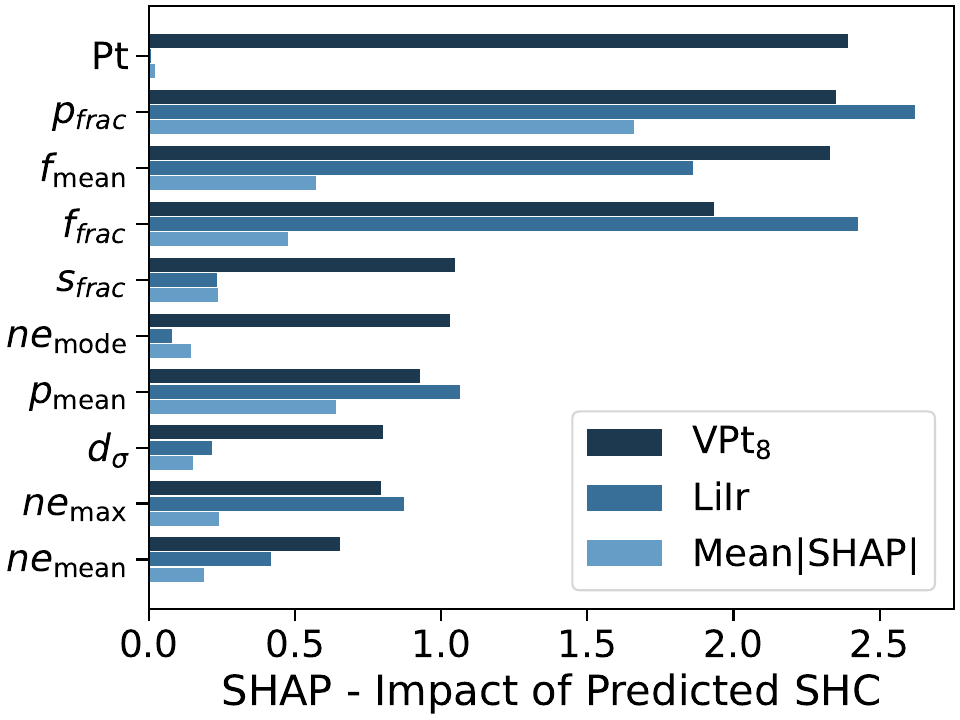}
		\caption{}
		\label{fig:shap_pos}
	\end{subfigure}
	\begin{subfigure}{0.49\textwidth}
		\includegraphics[width=\textwidth]{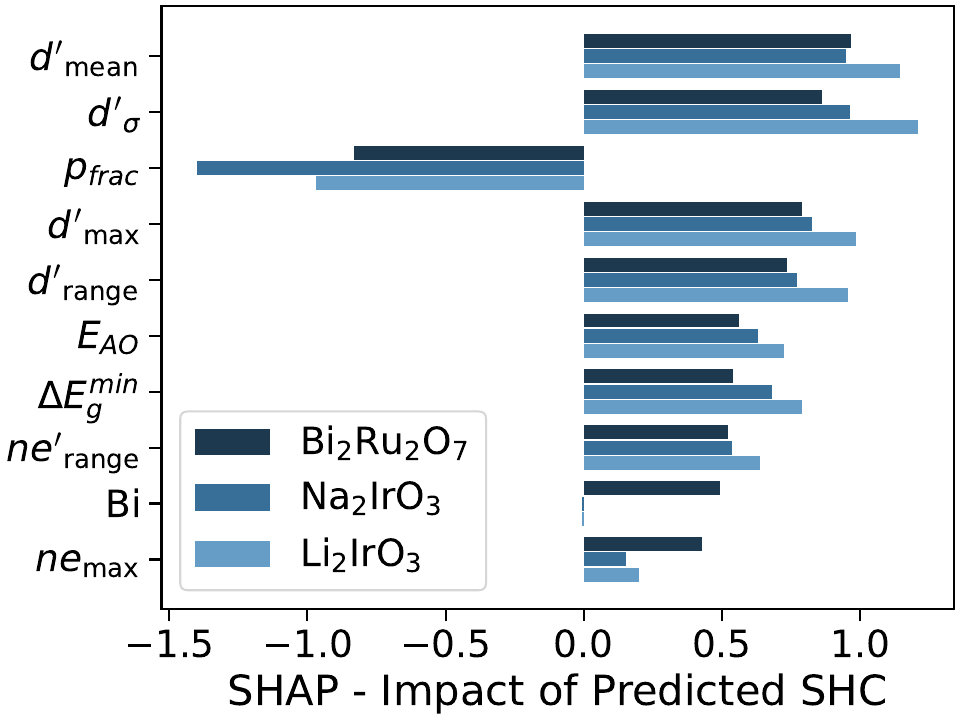}
		\caption{}
		\label{fig:shap_neg}
	\end{subfigure}
	\begin{subfigure}{0.32\textwidth}
		\includegraphics[width=\textwidth]{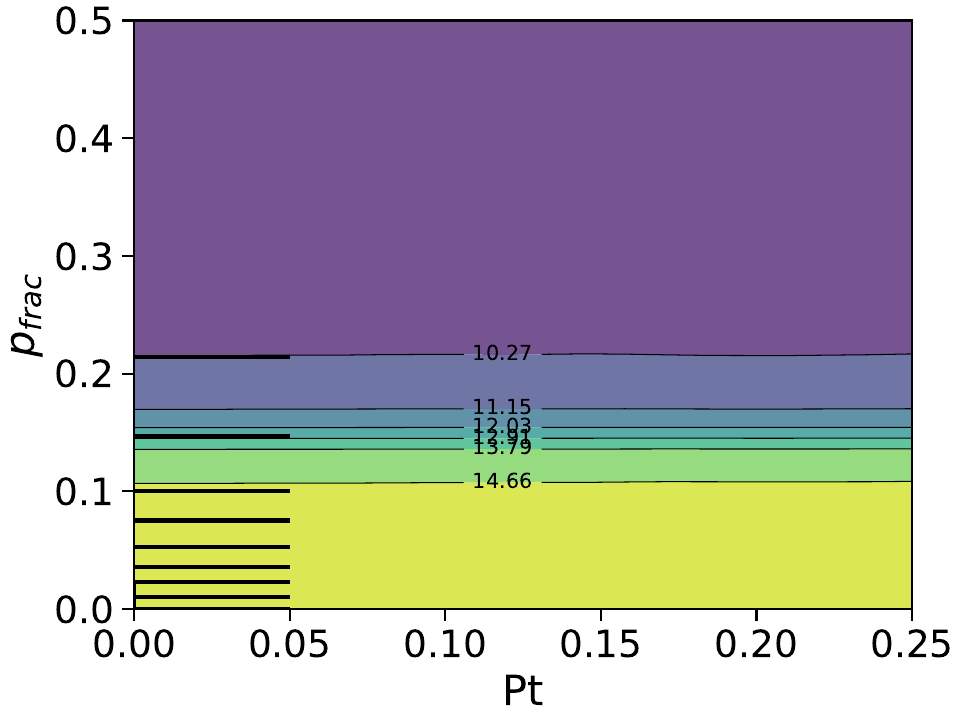}
		\caption{}
		\label{fig:pdp_pt}
	\end{subfigure}
	\begin{subfigure}{0.32\textwidth}
		\includegraphics[width=\textwidth]{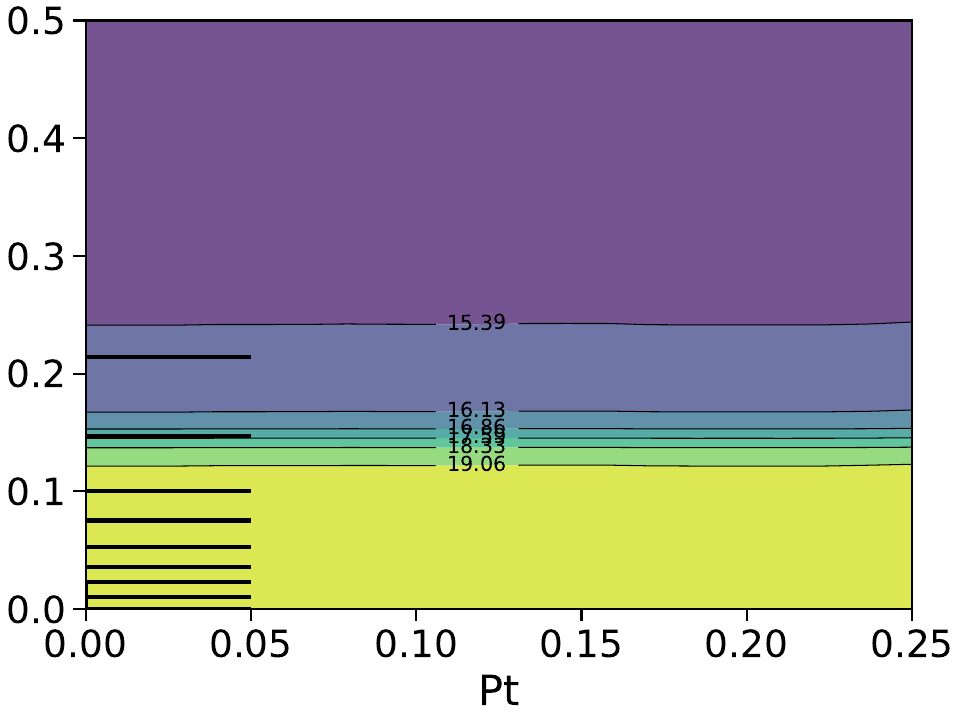}
		\caption{}
		\label{fig:pdp_Ta}
	\end{subfigure}
	\begin{subfigure}{0.32\textwidth}
		\includegraphics[width=\textwidth]{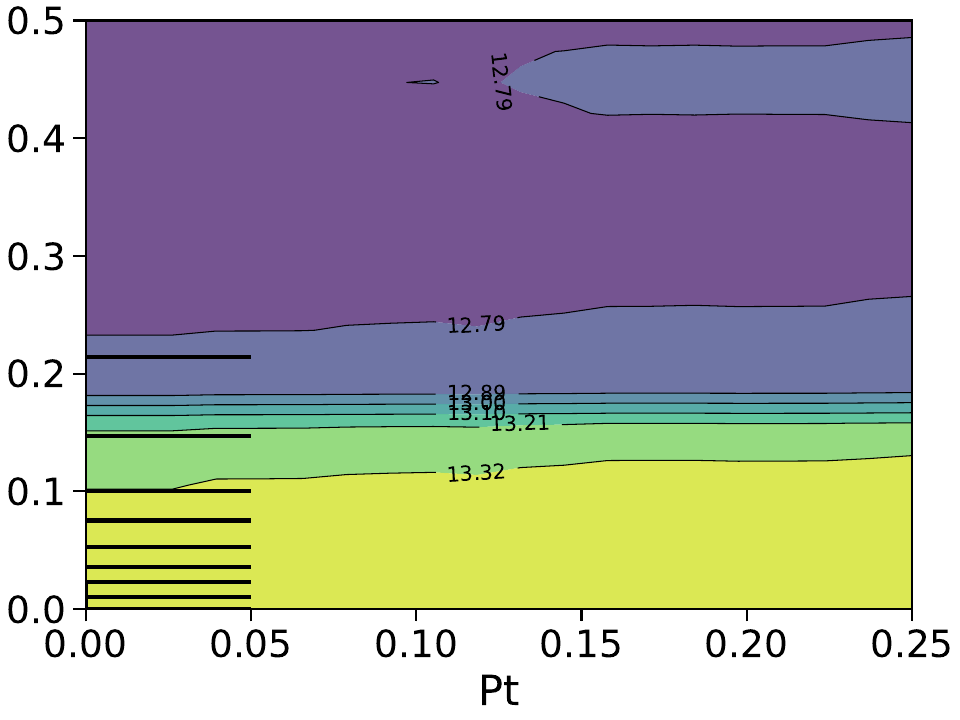}
		\caption{}
		\label{fig:pdp_Hf}
	\end{subfigure}
	\caption{Local SHAP explanations and counterfactual partial-dependence analysis. Figure (\subref{fig:shap_pos}) shows local
		contributions for \ce{VPt8} and \ce{LiIr}; figure (\subref{fig:shap_neg}) shows three instances where the model
		underestimates SHC by more than $900~(\hbar/e)(S/cm)$ on the training set - all Pt-free heavy-element compositions whose
		residual is dominated by the $p_{\text{frac}}$ and $\langle p \rangle$ features. Figure
		(\subref{fig:pdp_pt}) shows the effect of $p_{frac}$ under the presence of \ce{Pt}, the combined interaction shows the massive
		effect that $p_{frac}$ has on the predicted SHC. Figure (\subref{fig:pdp_Ta}) shows the partial-dependence on a subset
		containing heavy element compositions and figure (\subref{fig:pdp_Hf}) shows the same effect after deflating the statistical
		entanglement between the features \ce{Pt} and $p_{frac}$. Removing the dependence of $p_{frac}$ by counterfactual
		interrogation tones the effect down significantly with the gradient dropping from 3.7 (figure (\subref{fig:pdp_Ta})) to 0.53
		(figure (\subref{fig:pdp_Hf})) in boxcox-transformed units, illustrating that the model has learned to associate Pt content
		with enhanced SHC; held-out evaluation (\tabref{tab:screening}) shows that this association translates to a measurable but
		regime-dependent screening consequence.}
	%[comment: Changed figures 5(c) to 5(d) and added and appropriate caption.]
	%}
	\label{fig:local_shap}
\end{figure*}

The global attributions establish which features the model uses; the local explanations reveal how one of them is constructed,
and expose the sharper correlation--causation divergence anticipated above. \figref{fig:local_shap} presents local SHAP
explanations for representative high-SHC predictions. They reveal a dependency specific to the training distribution: because Pt
dominates the high-SHC entries of the Zhao~\textit{et~al.} dataset, the model has learned to tie the average $p$-valence
descriptor ($p_{\text{frac}}$) to the presence of Pt. Within this dataset $p_{\text{frac}}$ therefore functions partly as a
statistical marker for Pt content rather than purely as the proxy for $p$-orbital hybridization physics it was designed to be.
We stress that this is a statement about the model's learned representation against this dataset, not a physical claim that Pt is
uniquely necessary for high SHC - indeed, the first-principles verification below identifies Pt-free compounds with larger SHC
than any Pt compound in our set.

This effect is evident in (\figref{fig:pdp_pt}--\figref{fig:pdp_Hf}). Pt-containing compositions receive large positive SHAP
contributions from $p_{\text{frac}}$, while Pt-free heavy-element systems receive comparatively reduced contributions despite
hosting comparable atomic SOC strength. Counterfactual partial-dependence analysis on the dataset confirms the entanglement
directly: when the feature $p_{\text{frac}}$ is randomly permuted across instances containing \ce{Pt}, within the feature's
range, the entanglement is lifted synthetically and the average predicted SHC drops even though $p_{\text{frac}}$ spans the
same range (\figref{fig:pdp_Hf}).
This is a known failure mode for composition-only models when one element dominates a particular regime of the target
distribution\cite{breiman2001statistical}.

This entanglement is consistent across both the Random Forest and the Gaussian Process model. Two regressors with
qualitatively different inductive biases - axis-aligned splits versus a smooth RBF prior - converge on the same Pt-mediated
amplification. This Rashomon-style agreement\cite{semenova2022existence} strengthens the diagnosis: the entanglement is a
property of the compositional representation against the available training distribution, not of any one model class. It
cannot be cured by switching learners.

\subsection{An exploratory probe: does the entanglement have screening consequences?}

Having diagnosed the Pt-orbital entanglement, we ask whether it carries a practical consequence: does the model systematically
under-rank high-SHC Pt-free compositions, and if so, can a simple correction recover them? We frame this as an exploratory probe
of the diagnosis, not a validated screening method, and state the reason plainly - the correction factor is swept over a range
and evaluated on the same held-out set, so the analysis can establish \emph{whether} an exploitable signal exists but not a
correction that would generalize. We report it in that spirit.

The headline finding is that any benefit is small and confined to one regime. \tabref{tab:screening} reports recall at $k$ for
the threshold $y_{\text{high}} = 1000~(\hbar/e)(S/cm)$ under the reranking of \eqnref{eq:bias_aware}. Across the four thresholds
tested ($y_{\text{high}} \in \{500, 1000, 1500, 2000\}$) the correction helps in only one: at $y_{\text{high}} = 1000$ and
moderate top-$k$. At $y_{\text{high}} = 500$ it degrades recall at every $k \geq 100$; at $y_{\text{high}} = 1500$ and $2000$ both
protocols already achieve near-perfect recall ($n = 7$ and $n = 4$ true positives), so no discrimination is possible. Only in
the intermediate regime does a signal appear: bias-aware reranking with a minimal factor $\alpha_0 = 1.2$ raises recall from
$0.680$ to $0.760$ at $k = 150$ and from $0.800$ to $0.840$ at $k = 200$, robustly across $\alpha_0 \in [1.2, 1.6]$ and
reproduced by an additive correction $\beta = 200~(\hbar/e)(S/cm)$. The full sensitivity sweep is in the supplementary
information.

\begin{table}[t]
	\centering
	\caption{Recall at $k$ on the held-out test set ($N = 1503$, 25 true-high compounds with SHC $\geq 1000~(\hbar/e)(S/cm)$),
		for naive ranking and bias-aware reranking with three representative multiplicative correction factors $\alpha_0$.
		Bold entries indicate improvement over naive. Any improvement is confined to $k \geq 150$ at this threshold and does not
		appear at other thresholds (see text).}
	\resizebox{\columnwidth}{!}{
		\begin{tabular}{lccccc}
			\toprule
			Protocol                     & $k=20$  & $k=50$  & $k=100$ & $k=150$          & $k=200$          \\
			\midrule
			Naive                        & $0.240$ & $0.360$ & $0.600$ & $0.680$          & $0.800$          \\
			Bias-aware, $\alpha_0 = 1.2$ & $0.240$ & $0.360$ & $0.600$ & $\mathbf{0.760}$ & $\mathbf{0.840}$ \\
			Bias-aware, $\alpha_0 = 1.4$ & $0.200$ & $0.360$ & $0.600$ & $\mathbf{0.760}$ & $\mathbf{0.840}$ \\
			Bias-aware, $\alpha_0 = 1.6$ & $0.200$ & $0.360$ & $0.560$ & $\mathbf{0.760}$ & $\mathbf{0.840}$ \\
			\bottomrule
			\label{tab:screening}
		\end{tabular}
	}
\end{table}

The probe is best understood through the under-ranked compositions it recovers. The recall gain appears at $k = 150$ and
$k = 200$, where bias-aware reranking lifts one-to-two genuine high-SHC compounds across the top-$k$ boundary that naive
ranking leaves just below it; the clearest example is \ce{KBa2Re} (true SHC $1343~(\hbar/e)(S/cm)$ against an RF prediction of
$433$), promoted at $k = 100$, $\alpha_0 = 1.4$, although at that operating point its promotion is offset by the demotion of
another true-high so the recall is unchanged there. Its chemistry illustrates the diagnosis: \ce{KBa2Re} pairs the $5d$ element
Re with alkali and alkaline-earth partners, a combination absent from the Pt-dominated high-SHC tail, so its Berry-curvature
contribution arises through a hybridization pathway the model has not learned to associate with high SHC. The improvement
rests on recovering only a handful of such compounds, which is why we decline to present the correction as a method: it
correctly identifies genuinely under-ranked compositions, but it also promotes many low-SHC Pt-free compositions that a
multiplicative reweighting cannot distinguish from the true positives.

The probe therefore yields a qualified conclusion. The entanglement is real and has a detectable consequence - the model does
under-rank some high-SHC Pt-free compounds, and at least one is recoverable - but the consequence is too weak and too
regime-specific to support a screening correction. Recovering more would require conditioning variables richer than mere
Pt-absence (identifying the specific $5d$/$sp$/$f$/alkali pairings that generate intrinsic Berry curvature) or, more
fundamentally, training data with greater high-SHC Pt-free diversity - the same upstream remedy the audit points to
throughout. The lasting value of the probe is not the correction but the conversion it performs: a qualitative attribution
finding becomes a quantitative, falsifiable statement about the model's behavior on held-out data, exactly the kind of statement
that standard error metrics leave untested.

\subsection{Large-scale screening across Materials Project compositions}

\begin{figure}[t]
	\centering
	\includegraphics[width=\columnwidth]{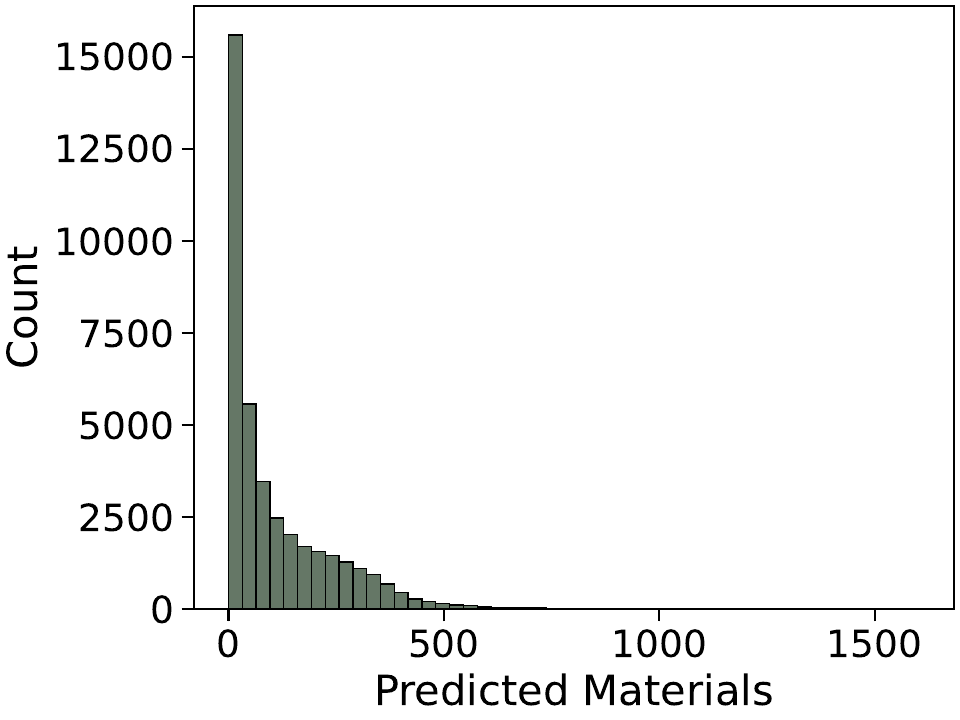}
	\caption{Distribution of Random-Forest-predicted SHC across the $\sim 40000$ screened Materials Project compositions. The
		predicted distribution reproduces the heavy right-skew of the training data, with the great majority of compositions
		predicted to have modest SHC and a sparse high-SHC tail forming the screening shortlist.}
	\label{fig:pred_count}
\end{figure}

Having characterized where the model is reliable and where it is not, we deploy it at scale. Using the trained Random Forest, we
performed inference over approximately 40000 inorganic compositions drawn from the Materials Project database - a regime far
beyond the reach of direct Kubo-Wannier evaluation, and one accessible to our model precisely because it requires no relaxed
structure as input. The predicted SHC distribution (\figref{fig:pred_count}) mirrors the right-skew of the training data: the
large majority of compositions are predicted to have modest SHC, with a sparse high-SHC tail that constitutes the screening
shortlist. Of the full set, 518 compositions are predicted to exceed $500~(\hbar/e)(S/cm)$, 15 exceed
$1000~(\hbar/e)(S/cm)$, and none exceed $2000~(\hbar/e)(S/cm)$, reducing forty thousand candidates to a tractable set for
first-principles follow-up. These counts are conservative by construction: the same Pt-orbital entanglement diagnosed above
causes the model to under-predict high-SHC compounds in Pt-free chemistries, so the true number of high-SHC compositions in this
pool is expected to exceed the model's flagged count --- as the DFT verification below confirms for \ce{HgOsPb2}, whose true SHC
of $2703~(\hbar/e)(S/cm)$ places it above a threshold its prediction of $717$ does not reach.

As an external check on the screening, \tabref{tab:reported_predictions} compares model predictions against literature SHC
values for compounds with available reference data, spanning elemental metals, A15 superconductors, topological insulators,
semimetals, and Heusler magnets. The model reproduces the correct order of magnitude and the relative ordering across these
chemically diverse families, supporting its use as a coarse filter - with the important qualification, developed next, that
its reliability is not uniform across chemistries.

\begin{figure*}[ht]
	\centering
	\begin{subfigure}[top]{0.6\textwidth}
		\includegraphics[width=\textwidth]{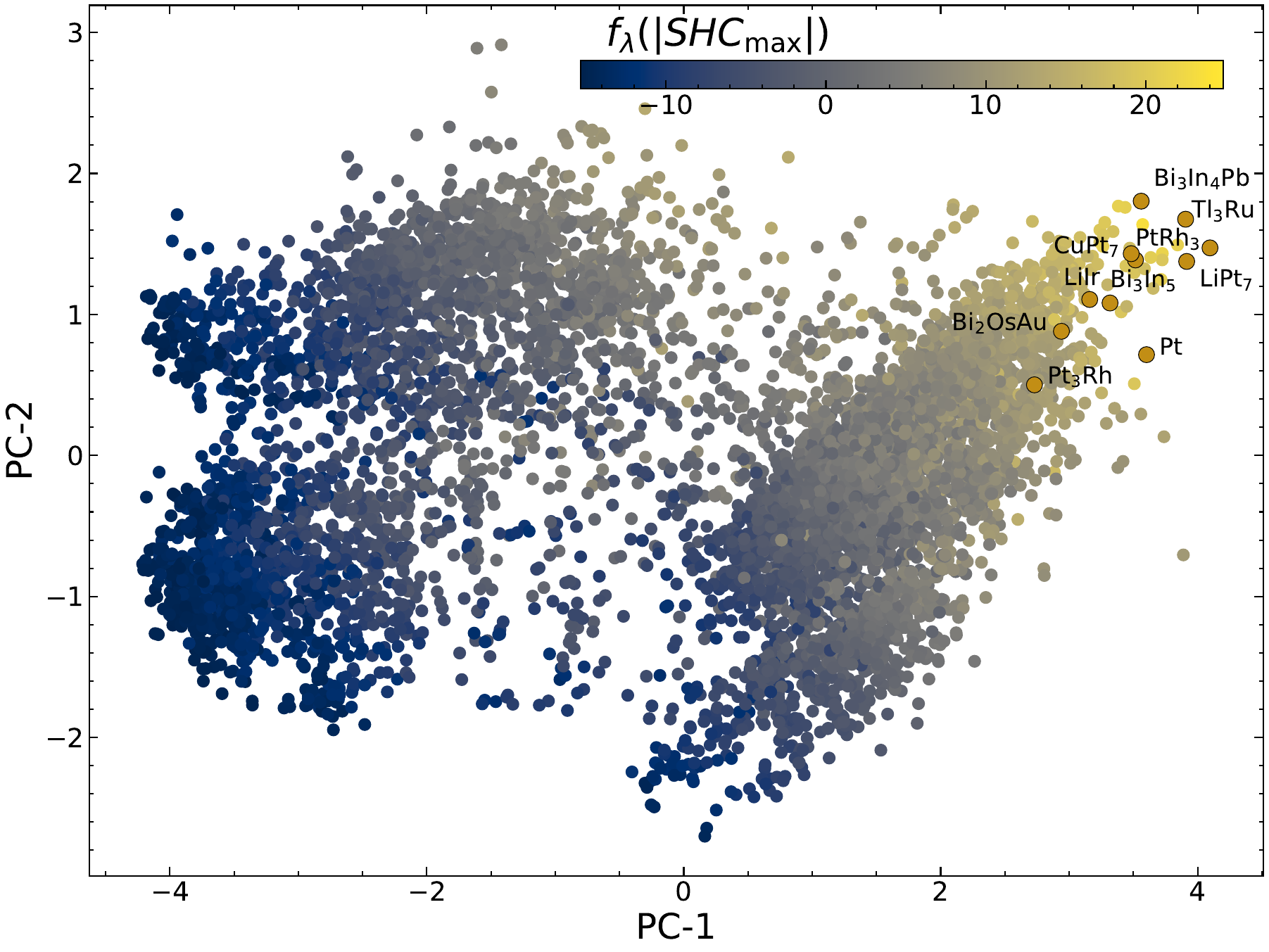}
		\caption{}
		\label{fig:shap_pca}
	\end{subfigure}
	\hfill
	\begin{minipage}[top]{.35\textwidth}
		\centering
		\begin{subfigure}{0.78\textwidth}
			\includegraphics[width=\textwidth]{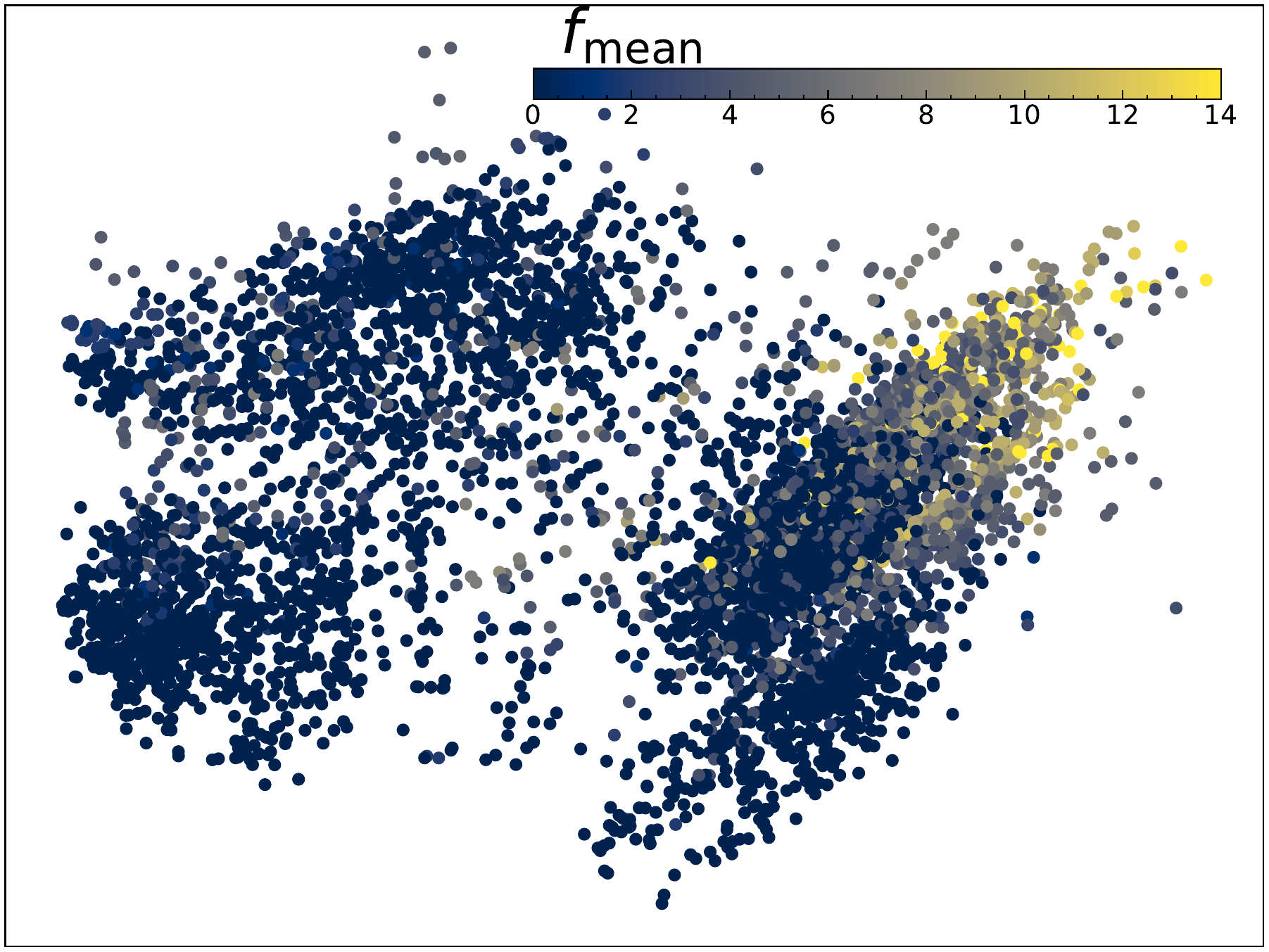}
			\caption{}
			\label{fig:shap_f_mean}
		\end{subfigure}
		\\[1ex]
		\begin{subfigure}{0.78\textwidth}
			\includegraphics[width=\textwidth]{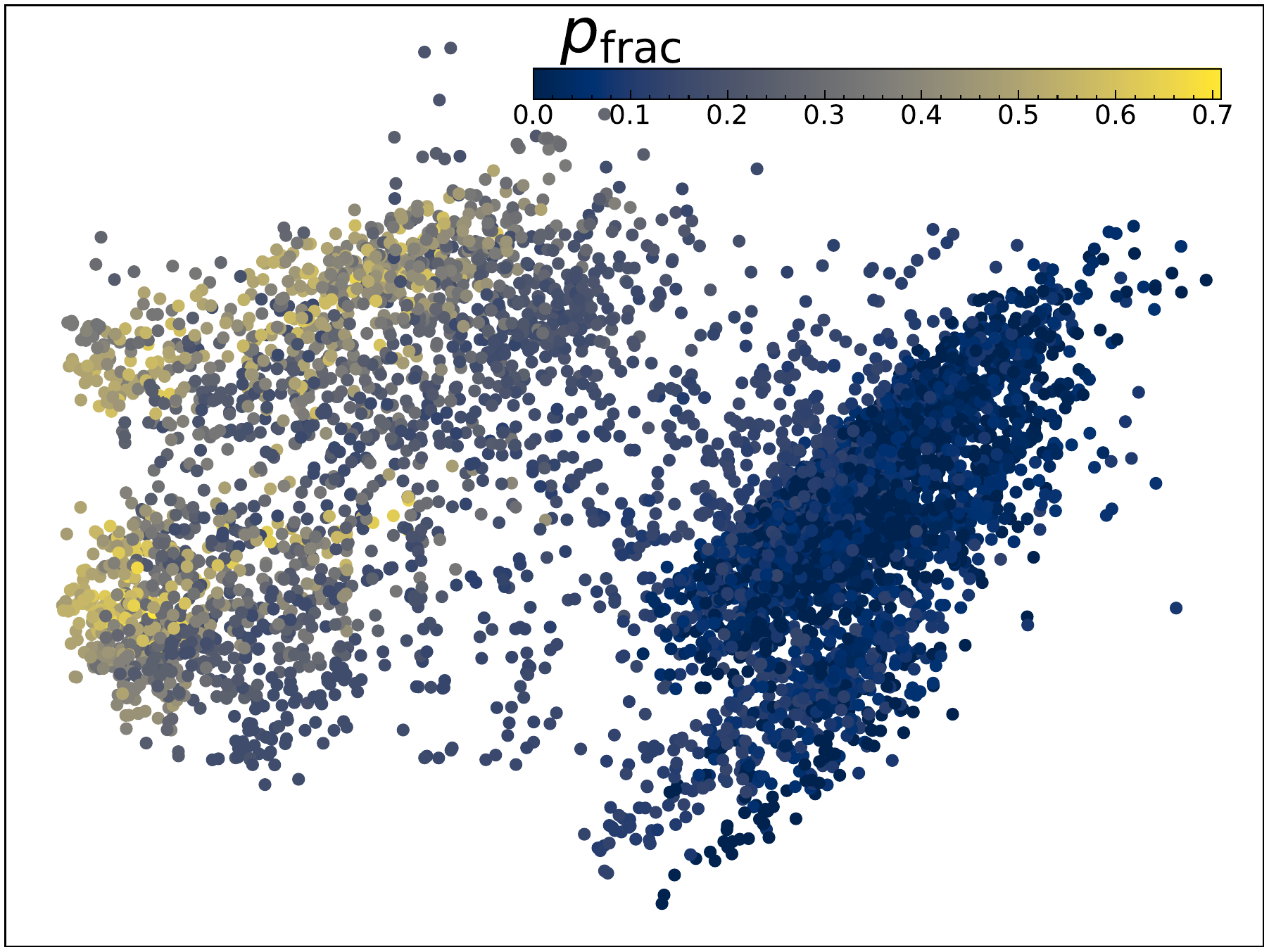}
			\caption{}
			\label{fig:shap_p_frac}
		\end{subfigure}
	\end{minipage}
	\caption{Figure (\subref{fig:shap_pca}) shows the two most significant principal components of the SHAP-transformed dataset.
		Materials with reported SHC values above $2000~(\hbar/e)(S/cm)$ (orange dots) cluster in a distinct region of the SHAP-PC
		space, indicating that the model has learned a coherent compositional pattern characteristic of high-SHC entries in the
		training set. Figures (\subref{fig:shap_f_mean}) and (\subref{fig:shap_p_frac}) decompose the same projection by the
		$\langle f \rangle$ and $p_{\text{frac}}$ feature values, respectively, showing that the cluster is driven by these
		two features in combination. SHC in figure (\subref{fig:shap_pca}) is in Box-Cox-transformed units.}
	\label{fig:pca}
\end{figure*}

The screening shortlist is dominated by exactly the regime the audit flagged as least reliable. Of the 7855 screened
compositions containing at least one heavy element from the set
$\{\text{Pt}, \text{Ta}, \text{W}, \text{Ir}, \text{Au}, \text{Bi}, \text{Hg}, \text{Hf}, \text{Re}, \text{Os}, \text{Pd}\}$,
only 1002 ($\approx 12.7\%$) contain Pt; the remaining 6853 ($\approx 87.3\%$) are Pt-free. The model's predictions for this
Pt-free majority carry the caveat established above: although the point predictions are statistically calibrated against the
training distribution, the model discriminates high- from low-SHC compositions less sharply here than within the Pt-containing
subset, because $p_{\text{frac}}$ - the descriptor it most relies on for high-SHC predictions - is entangled with Pt
content. The SHAP-PCA projection (\figref{fig:shap_pca}) confirms that high-SHC training instances occupy a distinct region of
feature space, so the model does recognize high-SHC \emph{patterns}; but Pt-free candidates near the boundary of that region are
the ones for which the predicted point estimate is least trustworthy, and for which independent verification is most valuable.
The first-principles calculations in the next section target precisely these cases, alongside agreement controls and a
deliberately chosen outlier.

\subsection{First-principles verification of the audit findings}

\begin{figure*}[ht]
	\centering
	\begin{subfigure}{0.24\textwidth}
		\includegraphics[width=\textwidth]{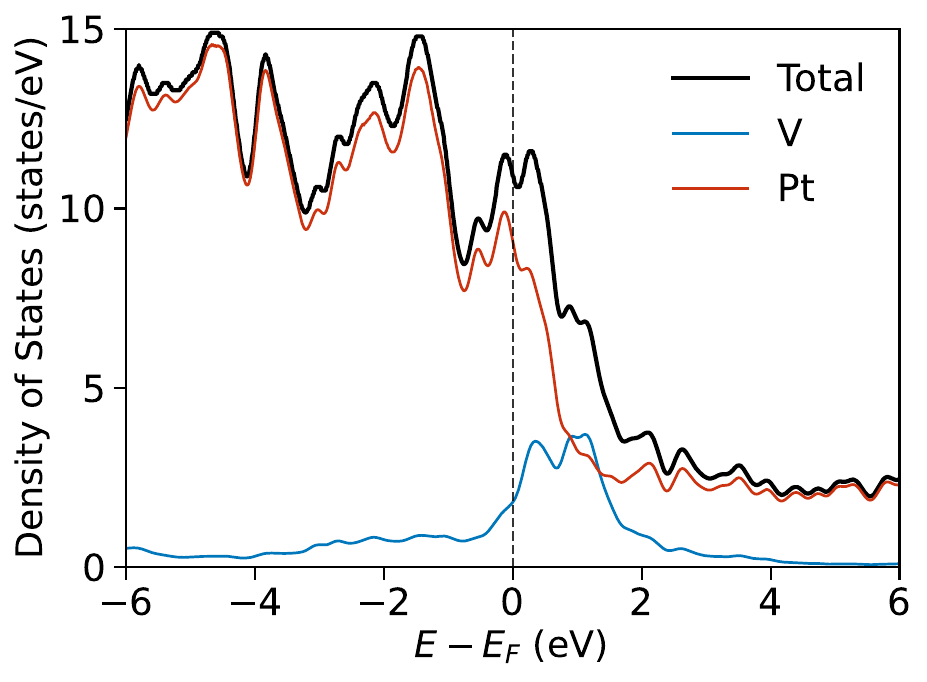}
		\caption{}
		\label{fig:pdos_vpt8}
	\end{subfigure}
	\begin{subfigure}{0.24\textwidth}
		\includegraphics[width=\textwidth]{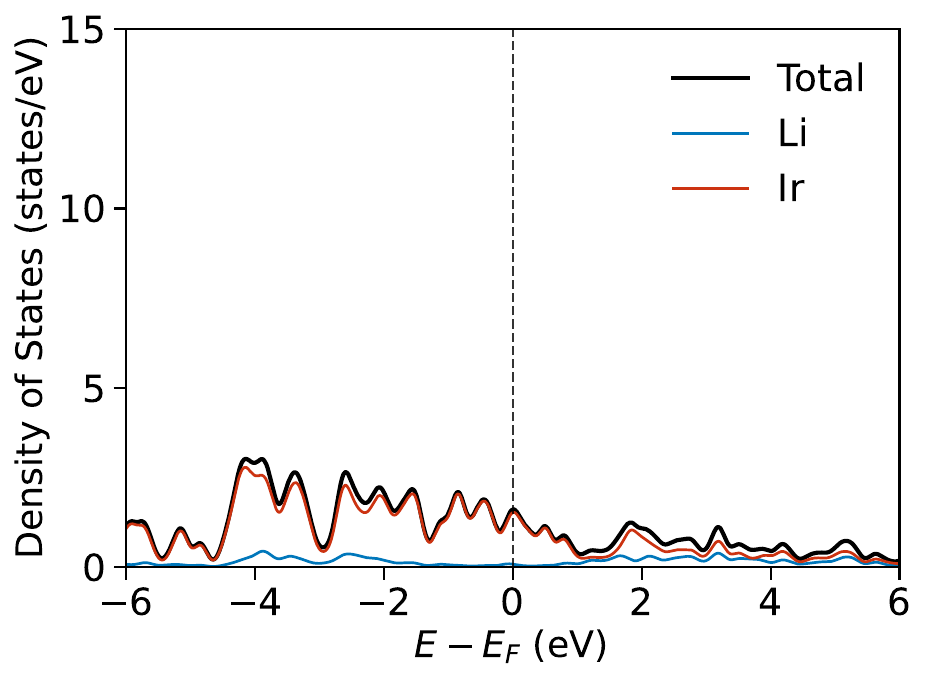}
		\caption{}
		\label{fig:pdos_liir}
	\end{subfigure}
	\begin{subfigure}{0.24\textwidth}
		\includegraphics[width=\textwidth]{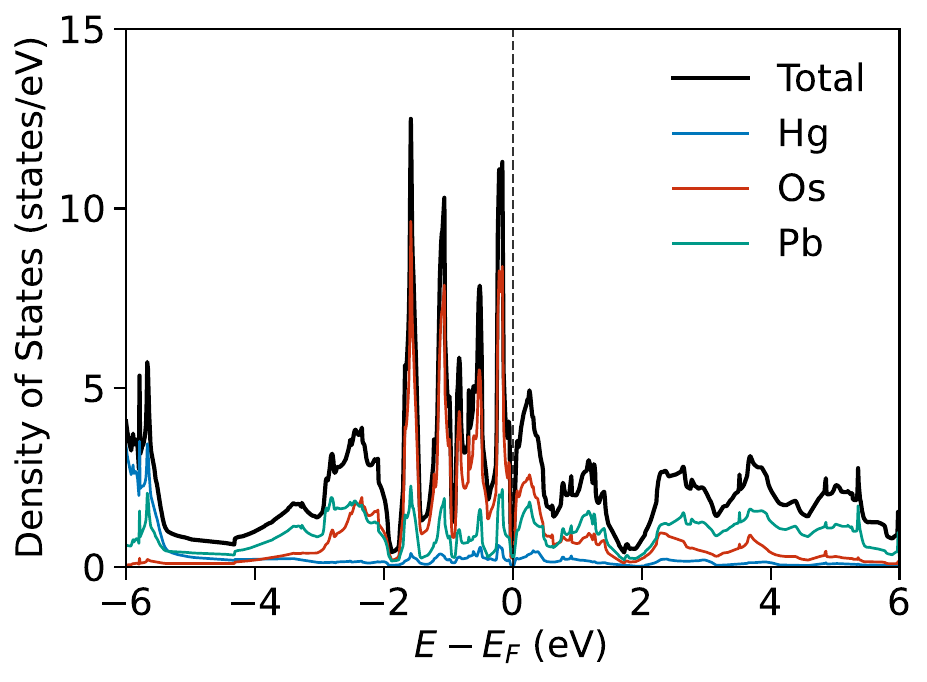}
		\caption{}
		\label{fig:pdos_hgospb2}
	\end{subfigure}
	\begin{subfigure}{0.24\textwidth}
		\includegraphics[width=\textwidth]{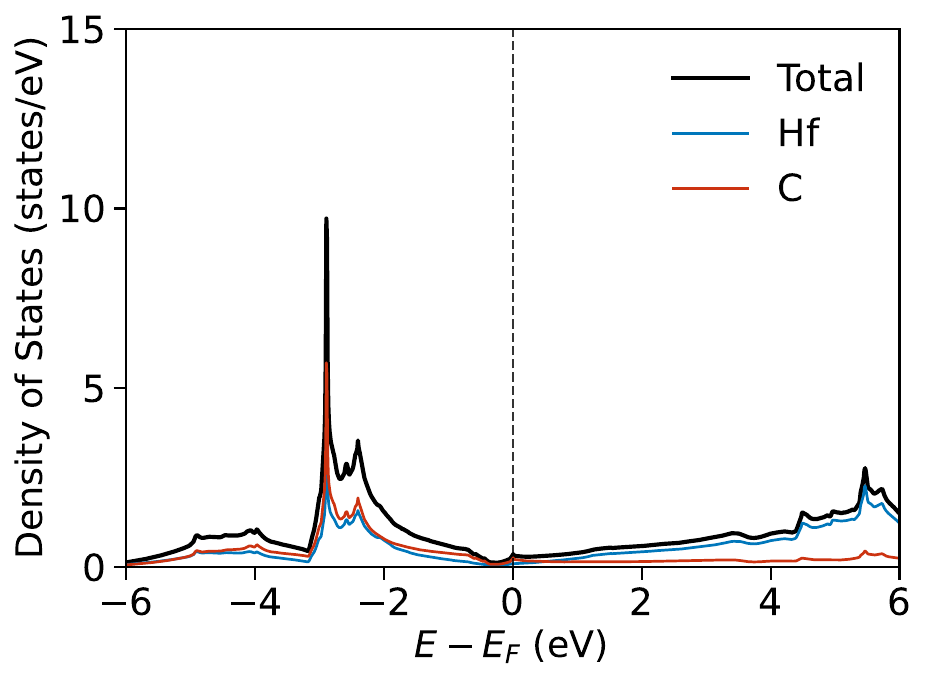}
		\caption{}
		\label{fig:pdos_hfc}
	\end{subfigure}

	\caption{Orbital-projected density of states (PDOS) for the four heavy-element validation compounds, with energy referenced
		to the Fermi level. (\subref{fig:pdos_vpt8}) \ce{VPt8}: dense Pt~$5d$ manifold at $E_F$ drives the confirmed high SHC.
		(\subref{fig:pdos_liir}) \ce{LiIr}: Ir~$5d$ states dominate while Li is an electronic spectator.
		(\subref{fig:pdos_hgospb2}) \ce{HgOsPb2}: Os~$5d$ states dominate at $E_F$, the origin of the DFT-confirmed
		$2703~(\hbar/e)(S/cm)$ peak that the composition-only model under-predicts by $\sim 3.8\times$.
		(\subref{fig:pdos_hfc}) \ce{HfC}: a low total DOS at $E_F$ with no dense $d$-manifold explains the small DFT SHC and the
		model's over-prediction. Energy window shown is $\pm 6$~eV about $E_F$.}
	\label{fig:pdos}
\end{figure*}

The audit findings above are hypotheses until independently tested. We therefore performed first-principles SHC calculations on
a set of compounds chosen to verify each finding: the model-audit prediction that high-SHC Pt-free compounds are under-flagged,
the data-audit hypothesis that a conspicuous model--label disagreement marks a faulty training label, and agreement controls
establishing the model's reliability where the audit raises no flag. For each compound we report two SHC measures: the peak
absolute component of $\sigma_{xy}^{z}(\varepsilon)$ over the full energy window, which matches the training-label convention of
Zhao~\textit{et~al.} and is used for the model--DFT comparison, and the maximum within $\pm 0.5$~eV of the Fermi level,
which is the physically relevant value for room-temperature spintronic operation. Projected densities of states for the four
heavy-element compounds are shown in \figref{fig:pdos}, and all DFT-versus-model values are collected in
\tabref{tab:dft_validation}.

\paragraph{Confirmed high-SHC discoveries (\ce{VPt8}, \ce{LiIr}).}
Both compounds were flagged by the model as high-SHC candidates and are confirmed by DFT. For \ce{VPt8}, the model predicted
$1452~(\hbar/e)(S/cm)$ and DFT yields a peak of $1393~(\hbar/e)(S/cm)$ - agreement within 4\%. The PDOS (\figref{fig:pdos_vpt8})
is dominated by the dense Pt~$5d$ manifold at $E_F$ (Pt contributing $\sim 14$~states/eV against $\sim 4$ for V), consistent
with hybridization between the Pt~$d_{5/2}$ states and V~$d$ states producing a broad, multi-band Berry-curvature distribution
and an SHC profile robust to chemical-potential shifts. For \ce{LiIr}, the model predicted $1141~(\hbar/e)(S/cm)$ while DFT
yields a substantially larger peak of $1826~(\hbar/e)(S/cm)$; the model correctly ranks \ce{LiIr} as high-SHC but
under-predicts its magnitude by $\sim 60\%$. The electronic profile (\figref{fig:pdos_liir}) is governed entirely by the
Ir~$5d$ manifold, with Li acting as an electronic spectator; strong atomic SOC splits the Ir~$d_{3/2}$ and $d_{5/2}$ states and
concentrates spin Berry curvature into a sharp resonant SHC peak near $E_F$.

\paragraph{The model audit verified: under-flagging of a Pt-free high-SHC compound (\ce{HgOsPb2}).}
This is the most direct test of the model-audit finding. \ce{HgOsPb2} is a Pt-free heavy-element composition that the model
predicts at only $717~(\hbar/e)(S/cm)$. Independent DFT yields a peak of $2703~(\hbar/e)(S/cm)$ - a factor of $3.8$ above the
prediction, and among the largest intrinsic SHC values reported for any nonmagnetic compound. The PDOS (\figref{fig:pdos_hgospb2})
shows that the states near $E_F$ are overwhelmingly of Os~$5d$ character (Os contributing $\sim 9.6$~states/eV against $0.6$ for
Hg and $2.3$ for Pb): the SHC originates from the Os~$5d$ manifold through a hybridization pathway distinct from the Pt~$d_{5/2}$
mechanism the model has learned to associate with high SHC. This compound is the empirical realization of the failure mode
diagnosed by SHAP - a genuinely high-SHC Pt-free material that the composition-only model systematically under-flags. We note
that \ce{HgOsPb2} lies $0.86$~eV/atom above the Materials Project convex hull and is therefore likely difficult to synthesize;
the DFT SHC is nonetheless a well-defined property of the relaxed structure and the under-prediction is real.

\paragraph{The data audit: a probable training-label error exposed (\ce{HfC}).}
\ce{HfC} demonstrates the second pillar of the protocol - auditing the training data itself. The audit was triggered by a
conspicuous model--label disagreement: the model predicts $695~(\hbar/e)(S/cm)$, presumably driven by the heavy Hf constituent,
while the Zhao~\textit{et~al.} training label is only $3.62~(\hbar/e)(S/cm)$, a discrepancy of nearly $200\times$ that stands
out even against the model's typical errors. Such an outlier admits two explanations - a severe model failure or a faulty label
- and only an independent calculation can distinguish them. Our DFT yields $120~(\hbar/e)(S/cm)$: the model \emph{is}
over-predicting, by a factor of $\sim 5.8$, but the training label is in error by a factor of $\sim 30$. The PDOS
(\figref{fig:pdos_hfc}) explains both findings at once. \ce{HfC} has a very low total density of states at $E_F$
($\sim 0.33$~states/eV), with Hf and C contributing comparably and modestly, and no dense $d$-manifold near the Fermi level to
host Berry curvature: composition signals strong SOC through Hf, which misleads the model upward, while the actual electronic
structure supports only a modest SHC - larger, however, than the near-zero dataset value.

The broader significance is that this error is invisible to every model trained on the dataset. CGCNN, Res-CGCNN, and the
SHCTransformer\cite{zhao2024accelerating, zhang2025predicting} all learn from the same labels and inherit the same error
without any mechanism to flag it; a black-box predictor that fits the faulty label well is rewarded, not corrected. The audit
protocol inverts this logic: a large, attribution-explicable model--label disagreement is treated not as a model failure to be
regularized away but as a hypothesis about the data, to be settled by targeted first-principles calculation at a cost of one
DFT run. To our knowledge this is the first demonstration, for a transport property, that explainability plus selective DFT
verification can perform quality control on a published training set.

\paragraph{Agreement controls (\ce{BiPt}, \ce{K2Pb2O3}).}
Two compounds from the held-out test set serve as agreement controls spanning the high- and low-SHC regimes. For the Pt-containing
\ce{BiPt}, the model predicts $814~(\hbar/e)(S/cm)$; DFT yields a near-Fermi value of $883~(\hbar/e)(S/cm)$ (full-window peak
$1229~(\hbar/e)(S/cm)$), an agreement within 8\% on the physically relevant measure. For the low-SHC \ce{K2Pb2O3}, the model
predicts $11~(\hbar/e)(S/cm)$ and DFT confirms a small SHC - a full-window peak of $84.6$ and a near-Fermi value of
$2.5$~$(\hbar/e)(S/cm)$ - placing the compound firmly in the low-SHC regime and confirming that the model correctly
identifies compositions with negligible spin Hall response.

\paragraph{Diagnostic under-prediction case (\ce{W3Ta}).}
\ce{W3Ta} is an A15-structure Pt-free compound for which the model predicts $770~(\hbar/e)(S/cm)$. Reference SHC values for this
compound differ between sources: the Zhao~\textit{et~al.} dataset lists $1011~(\hbar/e)(S/cm)$, while the dedicated A15
first-principles study of Refs.~\cite{derunova2019giant, sattigeri2024dirac} reports a substantially larger value of
$2250~(\hbar/e)(S/cm)$ arising from symmetry-demanded Dirac crossings near $E_F$. Against either reference, the composition-only
model under-predicts the SHC of \ce{W3Ta} - modestly relative to the dataset value and severely relative to the dedicated A15
calculation - consistent with the under-prediction trend for Pt-free heavy-element systems. We report \ce{W3Ta} as a
diagnostic case using literature values and do not perform an independent calculation here.

\paragraph{Summary.} The verification confirms both pillars of the audit. The model audit is verified: the model is accurate for
Pt-rich high-SHC compounds (\ce{VPt8}), and correctly ranks but under-predicts Ir- and Os-based Pt-free high-SHC compounds
(\ce{LiIr}, \ce{HgOsPb2}, \ce{W3Ta}), exactly the under-flagging the diagnosed Pt-orbital entanglement anticipates. The data
audit is verified: the conspicuous model--label disagreement on \ce{HfC} resolves into a probable training-label error, with the
model also over-predicting because composition signals SOC that the electronic structure does not realize. The agreement
controls (\ce{BiPt}, \ce{K2Pb2O3}) establish reliability where the audit raises no flag. The verification thus delineates,
through independent DFT, the precise chemical boundaries of the model's reliability while demonstrating that the protocol can
turn attribution-level findings into confirmed statements about both the model and its training data.

\begin{table*}[t]
	\centering
	\caption{First-principles validation of model predictions. SHC values are the maximum absolute component
		$|\sigma_{xy}^{z}|$; the full-window peak is compared against the model (consistent with the training-label convention),
		and the near-$E_F$ value (within $\pm 0.5$~eV) is reported where it differs. ``Reference'' values are from the
		Zhao~\textit{et~al.} dataset or the cited literature. All SHC in $(\hbar/e)(S/cm)$.}
	\label{tab:dft_validation}
	\begin{tabular}{llrrrrl}
		\toprule
		Compound     & Regime              & RF pred. & GPR conf. & DFT (peak) & DFT (near-$E_F$) & Reference                                                         \\
		\midrule
		\ce{VPt8}    & Agreement (Pt-rich) & $1452$   & $0.87$    & $1393$     & $1393$           & -                                                                 \\
		\ce{LiIr}    & Under-prediction    & $1141$   & $0.89$    & $1826$     & $1826$           & -                                                                 \\
		\ce{HgOsPb2} & Under-prediction    & $717$    & $0.87$    & $2703$     & $2703$           & -                                                                 \\
		\ce{HfC}     & Over-prediction     & $695$    & $0.88$    & $120$      & $120$            & $3.62$ (Zhao)                                                     \\
		\ce{BiPt}    & Agreement (high)    & $814$    & $0.88$    & $1229$     & $883$            & -                                                                 \\
		\ce{K2Pb2O3} & Agreement (low)     & $11$     & $0.89$    & $84.6$     & $2.5$            & -                                                                 \\
		\ce{W3Ta}    & Under-prediction    & $770$    & -         & -          & -                & $1011$ (Zhao); $2250$\cite{derunova2019giant, sattigeri2024dirac} \\
		\bottomrule
	\end{tabular}
\end{table*}

\begin{table*}
	\begin{tabular}{lrrrrc}
		\toprule
		Formula                                      & RF (predicted) & GP (predicted) & GP variance & GP confidence & Reported SHC  \\
		\midrule
		\ce{Pt}\cite{sagasta2016tuning}              & 1603.5         & 834.5          & 16.92       & 0.864         & $>$1600       \\
		\ce{Ta}\cite{sagasta2018unveiling}           & 277.7          & 374.11         & 19.75       & 0.841         & $820 \pm 120$ \\
		\ce{NbAs}$^{A15}$                            & 273.2          & 219.23         & 12.71       & 0.90          & 330           \\
		\ce{Nb3Os}$^{A15}$                           & 438.4          & 325.05         & 13.44       & 0.89          & 460           \\
		\ce{Mn3Pt}\cite{zhu2024crystal}              & 359.3          & 382.25         & 13.74       & 0.89          & $\approx 275$ \\
		\ce{Ta3Sb}$^{A15}$                           & 1041           & 424.85         & 14.33       & 0.88          & 1400          \\
		\ce{Nb3Ge}$^{A15}$                           & 720.3          & 367.47         & 12.58       & 0.90          & 1691.4        \\
		\ce{TaP}\cite{sun2016giant}                  & 690            & 511.24         & 13.33       & 0.89          & 603           \\
		\ce{Cr3Ir}$^\dagger{A15}$                    & 389.4          & 427.67         & 13.47       & 0.89          & 1209          \\
		\ce{Nb3Al}$^{A15}$                           & 307.1          & 283.91         & 12.92       & 0.90          & 440           \\
		\ce{Nb3Au}$^{A15}$                           & 676.7          & 307.53         & 14.05       & 0.89          & 1060          \\
		\ce{Ta3Sn}$^{A15}$                           & 902.8          & 471.37         & 13.55       & 0.89          & 620           \\
		\ce{NbP}\cite{sun2016giant}                  & 215.1          & 189.68         & 12.92       & 0.90          & 135           \\
		\ce{Bi2Se3}\cite{farzaneh2020intrinsic}      & 97             & 211.35         & 12.89       & 0.90          & 294           \\
		\ce{Bi2Te3}\cite{farzaneh2020intrinsic}      & 253.9          & 278.47         & 12.82       & 0.89          & 436           \\
		\ce{Cr3Os}$^\dagger$$^{A15}$                 & 392.8          & 430.29         & 13.67       & 0.89          & 40            \\
		\ce{Ti3Pt}                                   & 260.8          & 214.59         & 12.65       & 0.90          & 330           \\
		\ce{Ta3Au}$^{A15}$                           & 612.4          & 612.59         & 15.13       & 0.88          & 870           \\
		\ce{Nb3Bi}$^{A15}$                           & 621.2          & 475.53         & 14.22       & 0.89          & 670           \\
		\ce{Nb3Sn}$^\dagger$                         & 657.8          & 373.99         & 12.57       & 0.90          & 983.1         \\
		\ce{V3Pt}$^\dagger$                          & 132.1          & 227.11         & 12.84       & 0.90          & 440           \\
		\ce{TaAs}\cite{sun2016giant}                 & 731.5          & 619.91         & 14.69       & 0.88          & 781           \\
		\ce{Sb2Se3}\cite{farzaneh2020intrinsic}      & 37.4           & 66.10          & 12.87       & 0.90          & 187           \\
		\ce{Sb2Te3}\cite{farzaneh2020intrinsic}      & 53.9           & 79.67          & 12.64       & 0.90          & 226           \\
		\ce{BiSb}\cite{csahin2015tunable}            & 283.3          & 550.58         & 14.84       & 0.88          & $>$100        \\
		\ce{PtTe2}\cite{xu2020high}                  & 119.4          & 89.66          & 13.03       & 0.90          & 100--1000     \\
		\ce{WTe2}\cite{zhou2018intrinsic}            & 19.3           & 121.10         & 13.53       & 0.89          & 204           \\
		\ce{W3Ta}$^{A15}$                            & 770.5          & 421.17         & 14.19       & 0.89          & 2250          \\
		\ce{Rh2MnAl}$^m$                             & 753.2          & 251.39         & 12.94       & 0.90          & 729.14        \\
		\ce{Cu2CoSn}$^m$                             & 249.7          & 343.67         & 13.04       & 0.90          & 578.53        \\
		\ce{Co2MnAl}$^m$                             & 406.6          & 390.63         & 13.08       & 0.90          & 694.82        \\
		\ce{Co2MnGa}$^m$                             & 404            & 406.80         & 12.94       & 0.90          & 633.13        \\
		\ce{MoTe2}$^\dagger$\cite{zhou2018intrinsic} & 57.5           & 155.63         & 12.46       & 0.90          & 361           \\
		\ce{ZnSe}\cite{stern2006current}             & 1.3            & 35.56          & 13.00       & 0.90          & 0.01          \\
		\bottomrule
	\end{tabular}
	\caption{RF and GPR predictions versus literature SHC values. The RF and GPR are independently trained predictive models;
		the GPR variance and confidence columns represent the GPR's predictive variance (in $f_\lambda(\sigma_{xy}^z)$ units) and
		the corresponding 1$\sigma$ confidence interval. Magnetic materials ($^m$) are from Ji~\textit{et~al.}\cite{ji2022spin};
		A15 superconductors ($^{A15}$) are from Refs.~\cite{derunova2019giant, sattigeri2024dirac}; topological insulators and
		semimetals are cited inline. The maximum absolute component of the SHC tensor is used throughout, consistent with the
		training data. Entries marked $^\dagger$ are cases where the model substantially underestimates the literature value;
		these are predominantly Pt-free compositions, consistent with the Pt-orbital entanglement diagnosed in the main text.}
	\label{tab:reported_predictions}
\end{table*}

\section{Conclusion}

The contribution of this work is an audit protocol for materials machine learning: explainability methods - SHAP attribution,
counterfactual partial-dependence analysis, and Rashomon-style cross-model verification - coupled to targeted first-principles
verification, used to interrogate both a model's learned representation and the training data it learns from. The two pillars,
a model audit and a data audit, are independent of the specific findings they happen to return here: the protocol is
model-agnostic, applies to any predictor whose inputs admit attribution, and would retain its function were the particular
dependencies it uncovered in this demonstration absent.

Demonstrated on intrinsic spin Hall conductivity, the audit returned one verified finding per pillar. The model audit found
that, within the Zhao~\textit{et~al.} dataset, the average $p$-valence descriptor becomes statistically entangled with Pt
content - a property of the learned representation, consistent across Random Forest and Gaussian Process models, and not a
physical claim about SHC - and independent DFT confirmed the predicted consequence: a Pt-free compound, \ce{HgOsPb2}, whose
true SHC of $2703~(\hbar/e)(S/cm)$ is nearly four times the model's prediction. The data audit, triggered by a conspicuous
model--label disagreement, exposed a probable error in the \ce{HfC} training label, which our independent calculation exceeds by
a factor of $\sim 30$ - an error silently inherited by every black-box model trained on the same dataset. The supporting
Random Forest, competitive with structure-aware baselines (test MAE $114.5$ against $118.7$ for Res-CGCNN) without any
structural input, is in this framing a means rather than an end: accurate enough that auditing its representation is worthwhile.

The wider lesson is that the standard quality metrics of materials machine learning - held-out error, cross-validation,
benchmark comparison - certify neither of the two assumptions on which every such model rests: that its features track physics
rather than accidents of the training set, and that its labels are correct. Both assumptions are testable, and this work shows
how: attribution analysis to interrogate the representation, and a handful of targeted first-principles calculations to
adjudicate what the attribution finds. The cost is modest and the conclusions are falsifiable. As machine-learning surrogates
are increasingly trusted to direct experimental and computational effort, the ability to audit not only a model's accuracy but
the integrity of its reasoning and its data becomes not a refinement but a prerequisite - and one that grows more pressing in
exactly the regime, common to curated materials datasets, where a single element dominates the high-property tail.

\section*{Data Availability}
The trained Random Forest and Gaussian Process models, the compositional feature set, and the full set of Random-Forest-predicted
SHC values for the $\sim 40000$ screened Materials Project compositions will be available  upon publication. The DFT input and
output files (relaxation, self-consistent field, and Wannier-interpolation calculations) for all first-principles-validated
compounds (\ce{VPt8}, \ce{LiIr}, \ce{HgOsPb2}, \ce{HfC}, \ce{BiPt}, \ce{K2Pb2O3}) are provided as part of the supplementary data.
The dataset of Zhao~\textit{et~al.}\cite{zhao2024accelerating} used for training is publicly
available as described in the original publication.

\bibliographystyle{unsrt}
\bibliography{biblio}
\end{document}